\documentclass[aps,prl,preprint,nofootinbib]{revtex4-1}
\usepackage{amsfonts}
\usepackage{amsmath}
\usepackage{amssymb}
\usepackage{graphicx}
\usepackage{setspace}
\usepackage{CJK}
\usepackage{moresize}
\usepackage[utf8]{inputenc}
\usepackage[english]{babel}

\newcommand{\begM}{\begin{multline}}
\newcommand{\eM}{\end{multline}}

\def\){\Big)}
\def\({\Big(}

\begin{document}

\title{Canonical Noether and the energy-momentum non-uniqueness problem in linearized gravity}
\author{Mark Robert Baker}
\email{mbaker66@uwo.ca}
\affiliation{Department of Physics and Astronomy, The University of Western Ontario, London, ON, Canada N6A 3K7}
\affiliation{The Rotman Institute of Philosophy, The University of Western Ontario, London, ON, Canada N6A 5B7}

\date{\today}

\begin{abstract}
Recent research has highlighted the non-uniqueness problem of energy–momentum tensors in linearized gravity; many different tensors are published in the literature, yet for particular calculations a unique expression is required. It has been shown that (A) none of these spin-2 energy–momentum tensors are gauge invariant and (B) the Noether and Hilbert energy–momentum tensors are not, in general, equivalent; therefore uniqueness criteria is difficult to specify. Conventional wisdom states that the various published energy–momentum tensors for linearized gravity can be derived from the canonical Noether energy–momentum tensor of spin-2 Fierz-Pauli theory by adding ad-hoc `improvement' terms (the divergence of a superpotential and terms proportional to the equations of motion), that these superpotentials are in some way unique or physically significant, and that this implies some meaningful connection to the Noether procedure. To explore this question of uniqueness, we consider the most general possible energy–momentum tensor for linearized gravity with free coefficients using the Fock method. We express this most general energy–momentum tensor as the canonical Noether tensor, supplemented by the divergence of a general superpotential plus all possible terms proportional to the equations of motion. We then derive systems of equations which we solve in order to prove several key results for spin-2 Fierz–Pauli theory, most notably that there are infinitely many conserved energy–momentum tensors derivable from the `improvement' method, and there are infinitely many conserved symmetric energy–momentum tensors that follow from specifying the Belinfante superpotential alone. This disproves several recent claims that the Belinfante tensor is uniquely associated to the Hilbert tensor in spin-2 Fierz–Pauli theory. We give two new energy–momentum tensors of this form. Most importantly, since there are infinitely many energy–momentum tensors of this form, no meaningful or unique connection to Noether's first theorem can be claimed by application of the canonical Noether `improvement' method.
\end{abstract}

\maketitle

\section{1) Motivation}

\normalsize

The energy-momentum tensor is a fundamental object for a physical field theory. In electrodynamics, the Lorentz force law and Poynting's theorem are both expressed by the divergence of the uniquely accepted physical energy-momentum tensor $T^{\mu\nu}$. Several energy-momentum tensors that represent the linearized gravitational field exist in the literature. This causes an ambiguity in which the choice of energy-momentum tensor will impact the results of a given calculation and the conservation laws for the model as a whole. Selecting a unique $T^{\mu\nu}$ is problematic because Magnano and Sokolowski \cite{magnano2002} showed that one cannot obtain a gauge invariant energy-momentum tensor for spin-2 Fierz-Pauli theory; an essential characteristic of the uniquely defined physical energy-momentum tensors for e.g. electrodynamics and Yang-Mills theory. As noted by Magnano and Sokolowski \cite{magnano2002}: 

\begin{quote}
{\it{``A gravitational energy–momentum tensor is highly desirable for a number of reasons. For instance, it is emphasized in}} \cite{babak1999} {\it{that such a genuinely local tensor is required for a detailed description of cosmological perturbations in the early universe. $\dots$ Furthermore, the metric stress tensor derived in}} \cite{babak1999} {\it{has a number of nice properties and according to the authors, their $T^{\mu\nu}$ is the correct energy–momentum tensor for the gravitational field. $\dots$ Applying a physically undeniable condition that the energy–momentum tensor should have the same gauge invariance as the field equations, we also conclude that this approach to gravity does not furnish a physically acceptable notion of gravitational energy density.''}} 
\end{quote}

\noindent The claim in \cite{babak1999} that the Hilbert (metric) energy-momentum tensor (\ref{FPhilbert}) is the correct physical expression seems to contradict the Hulse-Taylor 1993 Nobel prize in physics \cite{hulse1975,taylor1982}, who used the equations Peters and Mathews developed \cite{peters1963} from the linearized Landau-Lifshitz energy-momentum pseudotensor (\ref{LLEMT}) to model energy loss due to gravitational radiation of a binary pulsar system \cite{landau1971}; this observationally supported model uses a $T^{\mu\nu}$ which does not correspond to the Hilbert (metric) energy-momentum tensor. There exist many other energy-momentum pseudotensors for the gravitational field in general relativity (i.e. Weinberg \cite{weinberg1972}, Papapetrou \cite{papapetrou1948}, M$\ddot{o}$ller \cite{moller1958}, Bergmann-Thomson \cite{bergmann1953}, etc.) that can be linearized about the Minkowski background, further complicating the question as to which is physically significant in linearized gravity. Different energy-momentum tensors will be claimed to be the physical expression for spin-2 Fierz-Pauli theory in different publications, hence the need to address the non-uniqueness problem of energy-momentum tensors in linearized gravity. This was emphasized recently by Bi{\v{c}}{\'a}k and Schmidt \cite{bicak2016}, and to some degree our paper builds on their results.

Recent research has also proved that the Noether and Hilbert (metric) energy-momentum tensors are not, in general, equivalent \cite{baker2021}; that is, the physical energy-momentum tensor derived directly from Noether's first theorem \cite{blaschke2016} (symmetric, gauge invariant, conserved and trace-free) is not always equivalent to the Hilbert tensor in Minkowski spacetime for the same Lagrangian density. This further complicates the possibility of uniquely expressing an energy-momentum tensor for a physical theory. It is frequently asserted throughout the literature, however, that the canonical Noether energy-momentum tensor $T^{\mu\nu}_C$ (a non-symmetric, non-gauge invariant expression) derived using only the 4-parameter Poincar{\'e} translation, is the starting point for the derivation of various energy-momentum tensors found by the ad-hoc addition of the divergence of superpotentials, and terms proportional to the equations of motion; these ad-hoc additions are often referred to as ``improvements''. This is true in linearized gravity, where different expressions for $T^{\mu\nu}$ are claimed to have a connection to Noether's first theorem due to these ad-hoc ``improvements'' of the canonical Noether energy-momentum tensor of spin-2 Fierz-Pauli theory. A good summary of some of these connections can be found in \cite{szabados1992}. We will briefly review the most common expressions (Hilbert and Landau-Lifshitz energy-momentum tensors) to highlight this point. We start from the differential identity following from Noether's theorem \cite{gelfand2012,kosmann2011,noether1918},

\begin{multline}
\left( \frac{\partial \mathcal{L}}{\partial \Phi_A}
 - \partial_\mu \frac{\partial \mathcal{L}}{\partial (\partial_\mu \Phi_A)} 
 + \partial_\mu \partial_\omega \frac{\partial \mathcal{L}}{\partial (\partial_\mu \partial_\omega \Phi_A)} + \dots \ \right) \delta \Phi_A
 \\
+ \partial_\mu \left(  \eta^{\mu\nu} \mathcal{L} \delta x_\nu
+ \frac{\partial \mathcal{L}}{\partial (\partial_\mu \Phi_A)} \delta \Phi_A 
+ \frac{\partial \mathcal{L}}{\partial (\partial_\mu \partial_\omega \Phi_A)} \partial_\omega \delta \Phi_A
- \left[ \partial_\omega \frac{\partial \mathcal{L}}{\partial (\partial_\mu \partial_\omega \Phi_A)} \right] \delta \Phi_A
+ ... \ \right) = 0 \ ,
   \label{genenergy}
\end{multline}

\noindent which is derived by asserting invariance of the action $S[\Phi_A (x_\alpha)]$ under infinitesimal changes in $\delta x_\nu$ and $\delta \Phi_A$. One can use the associated action symmetries of coordinates $\delta x_\nu$ and fields $\delta \Phi_A$ to derive on-shell conserved `Noether currents' $J^\mu$. The symmetries associated with the canonical Noether energy-momentum tensor are well known; they are the change in coordinates $\delta x_\nu = a_\nu$ (the 4-parameter translation of the 10 parameter Poincar{\'e} group), and transformation of fields $\delta \Phi_A = - (\partial^\nu \Phi_A) \delta x_\nu $. We begin with the spin-2 Fierz-Pauli Lagrangian density \cite{fierz1939},

\begin{equation}
\mathcal{L}_{FP} = \frac{1}{4} [\partial_\alpha h_\beta^\beta \partial^\alpha h_\gamma^\gamma 
- \partial_\alpha h_{\beta\gamma} \partial^\alpha h^{\beta\gamma} 
+ 2 \partial_\alpha h_{\beta\gamma} \partial^\gamma h^{\beta\alpha}
- 2 \partial^\alpha h_{\beta}^\beta \partial^\gamma h_{\gamma\alpha}] ,
\label{FPL}
\end{equation}

\noindent and the resulting spin-2 equation of motion $E^{\mu\nu}$ can be obtained from linearization of the Einstein tensor of general relativity. Equivalently, $E^{\mu\nu}$ follows from substitution of Equation (\ref{FPL}) into the Euler-Lagrange expression in Equation (\ref{genenergy}),

\begin{equation}
E^{\mu\nu} = \frac{1}{2}[- \eta^{\mu\nu}  \square h 
+  \square h^{\mu\nu} 
+   \partial^\mu  \partial^\nu h
-   \partial_\lambda  \partial^\nu h^{\mu\lambda}
-  \partial_\lambda   \partial^\mu h^{\nu\lambda}
+  \eta^{\mu\nu} \partial^\alpha   \partial^\beta h_{\alpha\beta}
 ] .
\label{FPEOM}
\end{equation}

To derive the canonical Noether energy-momentum tensor for a second rank $h_{\mu\nu}$, from Equation (\ref{genenergy}) with transformation of coordinates $\delta x_\nu = a_\nu$ (the 4-parameter translation), and the transformation of fields $\delta \Phi_A = - (\partial^\nu \Phi_A) \delta x_\nu $, we have the canonical Noether energy-momentum tensor for a second rank $h_{\mu\nu}$, namely $T^{\mu\nu}_C = \eta^{\mu\nu} \mathcal{L} 
- \frac{\partial \mathcal{L}}{\partial (\partial_\mu h_{\alpha\beta})} \partial^\nu h_{\alpha\beta} 
- \frac{\partial \mathcal{L}}{\partial (\partial_\mu \partial_\omega h_{\alpha\beta})} \partial_\omega \partial^\nu h_{\alpha\beta}
+ \left( \partial_\omega \frac{\partial \mathcal{L}}{\partial (\partial_\mu \partial_\omega h_{\alpha\beta})} \right) \partial^\nu h_{\alpha\beta}
+ \dots \ \ $. For spin-2 Fierz-Pauli theory, we have a Lagrangian density in Equation (\ref{FPL}) with terms of the form $\partial h \partial h$, thus the canonical Noether energy-momentum tensor reduces to $T^{\rho\sigma}_C =  \eta^{\rho\sigma} \mathcal{L}_{FP} 
- \frac{\partial \mathcal{L}_{FP}}{\partial (\partial_\rho h_{\mu\nu})} \partial^\sigma h_{\mu\nu}$, thus we obtain using Equation (\ref{FPL}) the canonical Noether energy-momentum tensor for spin-2 Fierz-Pauli theory,

\begin{multline}
T^{\rho\sigma}_C = \eta^{\rho\sigma} \mathcal{L}_{FP}
- \frac{1}{2} [   \partial^\rho h_\zeta^\zeta  \partial^\sigma h^\mu_{\mu}
-    \partial^\rho h^{\mu\nu}  \partial^\sigma h_{\mu\nu}
-  \partial^\nu h_{\zeta}^\zeta  \partial^\sigma h^\rho_{\nu}
-    \partial^\zeta h^\rho_{\zeta} \partial^\sigma h^\mu_{\mu}
+ 2 \partial^\mu h^{\nu\rho} \partial^\sigma h_{\mu\nu}
] . \label{FPcanon}
\end{multline}

\noindent On-shell conservation of the canonical Noether energy-momentum tensor is guaranteed by Noether's first theorem, which we can verify as $\partial_\rho T^{\rho\sigma}_C =  E^{\lambda\gamma} \partial^\sigma h_{\lambda\gamma}$. 

Using the Fierz-Pauli Lagrangian density in Equation (\ref{FPL}) we will now derive the Hilbert energy-momentum tensor. The Hilbert energy-momentum tensor for a classical gauge theory in Minkowski space is defined in as $T^{\gamma\rho}_H = \frac{2}{\sqrt{-g}} \frac{\delta \mathcal{L}}{\delta  g_{\gamma\rho}} \Big|_{g = \eta} $. The Hilbert energy-momentum tensor is derived from a Lagrangian density by replacing all ordinary derivatives with covariant derivatives $\partial \to \nabla$, replacing the Minkowski metric with the general metric tensor $\eta \to g$, and inserting the Jacobian term $\sqrt{-g}$. For spin-2 Fierz-Pauli theory we obtain the well known Hilbert energy-momentum tensor \cite{petrov2017},

\begin{multline}
  \hspace*{-1cm}
T^{\rho\sigma}_H = \frac{1}{4} \eta^{\rho\sigma}  [\partial_\alpha h_\beta^{\beta} \partial^\alpha h_\gamma^{\gamma} 
- \partial_\alpha h_{\beta\gamma} \partial^\alpha h^{\beta\gamma} 
+ 2 \partial_\alpha h_{\beta\gamma} \partial^\gamma h^{\beta\alpha}
+ 2   h^{\zeta \mu}  \partial_\zeta \partial_\mu h^\beta_{\beta}] 
- \partial^\rho h_{\beta\alpha} \partial^\alpha h^{\beta\sigma}
- \partial_\alpha h_\beta^{\rho} \partial^\sigma h^{\beta\alpha}
+ \partial_\alpha h^{\rho}_\beta \partial^\alpha h^{\sigma\beta}
\\
+  \partial_\zeta  h^{\rho\lambda } \partial_\lambda h^{\sigma\zeta} 
- \partial_\zeta  h^{\rho \sigma} \partial_\lambda h^{\lambda\zeta}
- \frac{1}{2}  \partial_\zeta h^{\rho \sigma}  \partial^\zeta h^\lambda_{\lambda}
- \frac{1}{2} \partial^\rho h_\beta^{\beta} \partial^\sigma h_\alpha^{\alpha} 
+ \frac{1}{2} \partial^\rho h_{\beta\alpha} \partial^\sigma h^{\beta\alpha} 
+ \frac{1}{2} \partial^\sigma h_\beta^{\beta} \partial^\alpha h_\alpha^{\rho}
+ \frac{1}{2} \partial^\rho h_\beta^{\beta} \partial^\alpha h_\alpha^{\sigma}
\\
+   h^{\rho\lambda }  \partial_\zeta \partial_\lambda h^{\sigma\zeta} 
+  h^{\sigma\lambda }   \partial_\zeta \partial_\lambda h^{\rho\zeta} 
- h^{\rho \sigma}  \partial_\zeta \partial_\lambda h^{\lambda\zeta}
- h^{\lambda \zeta}  \partial_\zeta  \partial_\lambda h^{\rho\sigma}
+ \frac{1}{2} h^{\rho \sigma}   \partial_\zeta \partial^\zeta h^\lambda_{\lambda}
- \frac{1}{2} h^{\rho \mu}  \partial^\sigma \partial_\mu h^\beta_{\beta}
- \frac{1}{2} h^{\sigma \mu}  \partial^\rho \partial_\mu h^\beta_{\beta} , \label{FPhilbert}
\end{multline}

\noindent which is conserved on-shell up to $\partial_\rho T^{\rho\sigma}_H = - 2 E^{\alpha\beta} \bar{\Gamma}^\sigma_{\alpha\beta} $, where $\bar{\Gamma}^\sigma_{\alpha\beta} = \frac{1}{2}( \partial_\alpha   h^{\sigma}_{\beta } +    \partial_\beta   h^\sigma_{\alpha } -    \partial^\sigma   h_{\beta \alpha})$ is the linearized Christoffel symbol of the second kind. Extracting (\ref{FPcanon}) from (\ref{FPhilbert}) we can write the Hilbert tensor (\ref{FPhilbert}) as the canonical tensor (\ref{FPcanon}) plus the divergence of a superpotential, and terms proportional to the equations of motion,

\begin{equation}
T^{\rho\sigma}_H = T^{\rho\sigma}_C + \partial_\gamma \Psi^{[\rho\gamma]\sigma}_H - 2 h^{\sigma}_{\beta} E^{\rho\beta} .
\label{converseresult}
\end{equation}

\noindent The Hilbert superpotential $\Psi^{[\rho\gamma]\sigma}_H = - \Psi^{[\gamma\rho]\sigma}_H$ found by rearranging (\ref{FPhilbert}) is,

\begin{multline}
\Psi^{[\rho\gamma]\sigma}_H =
 \frac{1}{2} \eta^{\rho\sigma} h^{\gamma\alpha} \partial_\alpha h_{\beta}^\beta 
- \frac{1}{2} \eta^{\gamma\sigma} h^{\rho\alpha} \partial_\alpha h_{\beta}^\beta
+ \frac{1}{2} h^{\gamma\sigma} \partial^\rho h_\beta^{\beta}
- \frac{1}{2}  h^{\rho \sigma}  \partial^\gamma h^\beta_{\beta}
\\
+ h^{\rho\lambda } \partial_\lambda h^{\gamma\sigma} 
-h^{\gamma\lambda }  \partial_\lambda h^{\rho\sigma}
+h^{\sigma\beta} \partial^\gamma h^{\rho}_\beta
-h^{\sigma\beta} \partial^\rho h^{\gamma}_\beta .
\label{conversesuperpotential}
\end{multline}

A superpotential $\Psi^{[\rho\gamma]\sigma}$ must be antisymmetric in two indices ($[\rho\gamma]$) because adding the divergence of a superpotential $\partial_\gamma \Psi^{[\rho\gamma]\sigma}$ to the canonical Noether expression must not affect the on-shell conservation ($\partial_\rho \partial_\gamma \Psi^{[\rho\gamma]\sigma} = 0$). This result was probably first noticed by Belinfante \cite{belinfante1940}. The Belinfante superpotential $b^{[\rho\gamma]\sigma}$ for spin-2 theory is exactly what is derived from the Hilbert superpotential $\Psi^{[\rho\gamma]\sigma}_H$, a result we will explore in detail; see for example (\ref{hilbeli}). The relationship between the Belinfante superpotential, canonical Noether energy-momentum tensor and Hilbert energy-momentum tensor is well established \cite{rosenfeld1940,gotay1992,babak1999,borokhov2002,saravi2004,forger2004,leclerc2006b,leclerc2006,pons2018}: the Hilbert energy-momentum tensor can be obtained by adding both the divergence of the Belinfante superpotential and specific terms proportional to the equations of motion to the canonical Noether energy-momentum tensor, which is consistent with the relationship in (\ref{converseresult}).

We will now introduce the linearized Landau-Lifshitz energy-momentum tensor \cite{bicak2016},

\begin{multline}
T^{\mu\nu}_{LL} = 
\frac{3}{4} \eta^{\mu\nu}  \partial_\alpha h  \partial^\alpha h 
- \eta^{\mu\nu}  \partial_\alpha h \partial_\beta h^{\alpha\beta} 
+  \frac{1}{2} \eta^{\mu\nu} \partial^\lambda h^{\alpha\gamma}  \partial_\gamma h_{\lambda\alpha} 
- \frac{1}{4} \eta^{\mu\nu}  \partial^\alpha h^{\lambda\sigma} \partial_\alpha h_{\lambda\sigma}
-     \partial^\mu h    \partial^\nu h
\\
- \frac{3}{2}  \partial_\alpha h^{\mu\nu}  \partial^\alpha h 
+ \partial_\alpha h^{\mu\nu} \partial_\beta h^{\alpha\beta}
- \partial_\alpha h^{\mu\alpha} \partial_\beta h^{\nu\beta}
+ \frac{1}{2}  \partial^\mu h^{\lambda\sigma} \partial^\nu h_{\lambda\sigma}
+  \partial_\lambda h^{\mu\alpha} \partial^\lambda h^{\nu}_{\ \alpha}
\\
+ (\partial_\alpha h^{\mu\alpha}    \partial^\nu h 
+   \partial_\alpha h^{\nu\alpha}   \partial^\mu h)
+ \frac{1}{2} ( \partial^\mu h^{\nu\gamma}  \partial_\gamma h
+ \partial^\nu h^{\mu\gamma}  \partial_\gamma h)
-  ( \partial^\mu h_{\beta\gamma} \partial^\gamma h^{\nu\beta}
+ \partial^\nu h_{\beta\gamma} \partial^\gamma h^{\mu\beta}) . \label{LLEMT}
\end{multline}

\noindent This energy-momentum tensor can also be expressed in terms of (\ref{FPcanon}) and terms proportional to the equations of motion (\ref{FPEOM}) as \cite{szabados1992},

\begin{equation}
T^{\mu\nu}_{LL} = 
T^{\mu\nu}_C
+  \partial_\alpha \Psi^{[\mu\alpha]\nu}_{LL}
+  h E^{\mu\nu}
- 2 h_\beta^{\nu} E^{\mu\beta} ,
\label{LLadhocimproved}
\end{equation}

\noindent where the Landau-Lifshitz superpotential is,

\begin{multline}
\Psi^{[\mu\alpha]\nu}_{LL} =
\frac{1}{2} [ \eta^{\mu\nu} h  \partial^\alpha h 
-     \eta^{\nu\alpha} h \partial^\mu h    
+  \eta^{\nu\alpha} h \partial_\beta h^{\mu\beta}  
-   \eta^{\mu\nu} h \partial_\beta h^{\alpha\beta} 
+  h \partial^\mu h^{\nu\alpha} 
-  h \partial^\alpha h^{\mu\nu} ]
\\
+  h^{\nu\alpha}   \partial^\mu h
-   h^{\mu\nu}  \partial^\alpha h 
+ h^{\mu\nu} \partial_\beta h^{\alpha\beta}
- h^{\nu\alpha} \partial_\lambda h^{\mu\lambda}
+  h^{\nu}_{\beta}  \partial^\alpha h^{\mu\beta}
-  h^{\nu}_{\beta} \partial^\mu h^{\beta\alpha} .
\label{LLknownsuper}
\end{multline}

Both the Hilbert $T^{\rho\sigma}_H$ and Landau-Lifshitz $T^{\mu\nu}_{LL}$ energy-momentum tensors can be obtained by starting from the canonical Noether energy-momentum tensor of spin-2 Fierz-Pauli theory $T^{\rho\sigma}_C$, then ad-hoc adding the divergence of a superpotential, and terms proportional to the equations of motion (\ref{FPEOM}). This is frequently used to assert that these results can in some way be derived from Noether's first theorem. For example, in the Padmanabhan-Deser debate, \cite{padmanabhan2008,butcher2009,deser2010,butcher2012,barcelo2014},
Padmanabhan asserted \cite{padmanabhan2008} that for self coupling of the spin-2 energy-momentum tensor, one can add infinitely many different superpotentials to the canonical Noether tensor of spin-2 Fierz-Pauli theory, thus Noether's procedure cannot be used to determine the energy-momentum tensor. Subsequent authors \cite{deser2010,barcelo2014}, asserted that Noether's theorem can be used by adding the Belinfante superpotential and additional terms proportional to the equations of motion to the canonical Noether tensor to obtain $T^{\rho\sigma}_H$ in (\ref{FPhilbert}). But this ad-hoc addition of terms can also be used to obtain other published expressions, such as the Landau-Lifshitz in $T^{\mu\nu}_{LL}$ (\ref{LLEMT}). Since known energy-momentum tensors can be obtained by adding the ad-hoc correction terms to the canonical Noether tensor, this ``improvement'' process is portrayed as a meaningful connection of any such energy-momentum tensor to Noether's first theorem. However, some honest discussion of these ad-hoc ``improvements'' can be found in the literature, such as statements made by Forger and R\"{o}mer \cite{forger2004}: 

\begin{quote}
{\it{``There is a long history of attempts to cure these diseases and arrive at the physically correct energy-momentum tensor $T^{\mu\nu}$ by adding judiciously chosen ‘‘improvement’’ terms to [$\ T^{\mu\nu}_C$]''}}. They go on to say: {\it{``However, all these methods of defining improved energy-momentum tensors are largely ‘‘ad hoc’’ prescriptions focussed on special models of field theory, often geared to the needs of quantum field theory and ungeometric in spirit''}}. 
\end{quote}

\noindent We point out that Bessel-Hagen (a contemporary and colleague of Noether) first showed how to derive the physical energy-momentum tensor directly from Noether's first theorem without the need for any such ad-hoc ``improvements'' in 1921 \cite{besselhagen1921}. This result was determined independently by later authors (\cite{eriksen1980,burgess2002,montesinos2006}, to name a few) and summarized in \cite{blaschke2016}. Furthermore, it has recently been shown that the Noether and Hilbert energy-momentum tensor are not, in general, equivalent \cite{baker2021}, which further emphasizes the need for the investigation into the relationship between tensors which are derived directly from Noether's first theorem, and those which can only be obtained after adding the divergence of a superpotential and terms proportional to the equations of motion for a particular theory. This will be a subject that we address in this article.

Bi{\v{c}}{\'a}k and Schmidt explored the non-uniqueness of the energy-momentum tensors in linearized gravity in a recent article \cite{bicak2016} (Bi{\v{c}}{\'a}k has long be interested in this question \cite{bicak1965}). They used the Fock method for deriving an energy-momentum tensor \cite{fock2015,infeld1964}, which considers general expressions of terms with free coefficients for $T^{\rho\sigma}$. In particular, Bi{\v{c}}{\'a}k and Schmidt consider all possible terms of the form $\partial h \partial h$; we will denote their Fock energy-momentum tensor as $T^{\rho\sigma}_{BS}$,

\begin{multline}
T^{\rho\sigma}_{BS} = 
 b_1 \partial_\alpha h^{\rho\sigma} \partial_\beta h^{\alpha\beta}
+ b_2 \partial_\alpha h^{\rho\sigma} \partial^\alpha h 
+ b_3 \partial_\alpha h^{\rho\alpha} \partial_\beta h^{\sigma\beta}
+ b_4 \partial_\alpha h^{\rho}_{\ \beta} \partial^\alpha h^{\sigma\beta}
+ b_5 \partial_\alpha h^{\rho\beta} \partial_\beta h^{\sigma\alpha}
\\
+ b_6 \partial^\rho h \partial^\sigma h 
+ b_7 \partial^\rho h_{\alpha\beta} \partial^\sigma h^{\alpha\beta}
+ b_{8_{i}} \partial^\rho h^{\sigma\alpha} \partial_\alpha h 
+ b_{8_{ii}} \partial^\sigma h^{\rho\alpha} \partial_\alpha h
+ b_{9_{i}} \partial^\rho h \partial_\alpha h^{\sigma\alpha} 
+ b_{9_{ii}} \partial^\sigma h \partial_\alpha h^{\rho\alpha}
\\
+ b_{10_{i}} \partial^\rho h^{\sigma\alpha} \partial^\beta h_{\alpha\beta} 
+ b_{10_{ii}} \partial^\sigma h^{\rho\alpha} \partial^\beta h_{\alpha\beta}
+ b_{11_{i}} \partial^\rho h_{\alpha\beta} \partial^\alpha h^{\sigma\beta} 
+ b_{11_{ii}} \partial^\sigma h_{\alpha\beta} \partial^\alpha h^{\rho\beta}
\\
+ c_1 \eta^{\rho\sigma} \partial_\alpha h \partial^\alpha h
+ c_2 \eta^{\rho\sigma} \partial_\alpha h_{\beta\lambda} \partial^\alpha h^{\beta\lambda}
+ c_3 \eta^{\rho\sigma} \partial_\alpha h^{\alpha\beta} \partial_\lambda h^{\lambda}_{\ \beta}
+ c_4 \eta^{\rho\sigma} \partial_\alpha h_{\lambda\beta} \partial^\lambda h^{\alpha\beta}
+ c_5 \eta^{\rho\sigma} \partial_\alpha h^{\alpha\beta} \partial_\beta h  . \label{BStensor}
\end{multline}

\noindent This Fock energy-momentum tensor appears in Equation 8 of their article \cite{bicak2016}. In (\ref{BStensor}) we separate terms proportional to the Minkowski metric $\eta^{\rho\sigma}$ with coefficients $c_n$. The authors use this to prove some very interesting results which we have also verified, such as the uniqueness of the Landau-Liftshitz tensor as the conserved and symmetric expression that follows from (\ref{BStensor}). The problem is that (\ref{BStensor}) is not the most general expression, because many conserved energy-momentum tensors have terms of the form $h \partial \partial h$, such as the Hilbert tensor (\ref{FPhilbert}). The appearance of such terms ($h \partial \partial h$) greatly complicates the resulting linear system of coefficients. To accommodate these additional terms we will take a similar approach to \cite{bicak2016}, but instead we consider the most general possible Fock energy-momentum tensor for linearized gravity. This will be used to complete several proofs regarding the energy-momentum tensors in linearized gravity. In particular, we will consider the most general system which can be obtained by adding the divergence of a superpotential and terms proportion to the equations of motion to the canonical Noether energy-momentum tensor $T^{\mu\nu}_C$. Using this expression we will prove that there are infinitely many conserved tensors that can be obtained by the ad-hoc addition of these terms to $T^{\mu\nu}_C$, and that there are infinitely many symmetric conserved energy-momentum tensors following from the Belinfante improvement procedure alone. We argue that these results show that no meaningful connection to Noether's first theorem exists from the superpotential approach; if a tensor is not directly derived from Noether's first theorem, then it simply is not derived from Noether's first theorem, and no amount of ad-hoc ``improvements'' can change this fact.

\section{2) The most general Fock energy-momentum tensor for linearized gravity}

Since Bi{\v{c}}{\'a}k and Schmidt in \cite{bicak2016} consider (\ref{BStensor}), which is not the most general energy-momentum tensor for linearized gravity, as it does not include e.g. the Hilbert energy-momentum tensor in (\ref{FPhilbert}). We will begin our proofs with the most general expression that also includes terms of the form $h\partial \partial h$,

\begin{multline}
T^{\rho\sigma} = 
 b_1 \partial_\alpha h^{\rho\sigma} \partial_\beta h^{\alpha\beta}
+ b_2 \partial_\alpha h^{\rho\sigma} \partial^\alpha h 
+ b_3 \partial_\alpha h^{\rho\alpha} \partial_\beta h^{\sigma\beta}
+ b_4 \partial_\alpha h^{\rho}_{\ \beta} \partial^\alpha h^{\sigma\beta}
+ b_5 \partial_\alpha h^{\rho\beta} \partial_\beta h^{\sigma\alpha}
\\
+ b_6 \partial^\rho h \partial^\sigma h 
+ b_7 \partial^\rho h_{\alpha\beta} \partial^\sigma h^{\alpha\beta}
+ b_{8_{i}} \partial^\rho h^{\sigma\alpha} \partial_\alpha h 
+ b_{8_{ii}} \partial^\sigma h^{\rho\alpha} \partial_\alpha h
+ b_{9_{i}} \partial^\rho h \partial_\alpha h^{\sigma\alpha} 
+ b_{9_{ii}} \partial^\sigma h \partial_\alpha h^{\rho\alpha}
\\
+ b_{10_{i}} \partial^\rho h^{\sigma\alpha} \partial^\beta h_{\alpha\beta} 
+ b_{10_{ii}} \partial^\sigma h^{\rho\alpha} \partial^\beta h_{\alpha\beta}
+ b_{11_{i}} \partial^\rho h_{\alpha\beta} \partial^\alpha h^{\sigma\beta} 
+ b_{11_{ii}} \partial^\sigma h_{\alpha\beta} \partial^\alpha h^{\rho\beta}
\\
+ c_1 \eta^{\rho\sigma} \partial_\alpha h \partial^\alpha h
+ c_2 \eta^{\rho\sigma} \partial_\alpha h_{\beta\lambda} \partial^\alpha h^{\beta\lambda}
+ c_3 \eta^{\rho\sigma} \partial_\alpha h^{\alpha\beta} \partial_\lambda h^{\lambda}_{\ \beta}
+ c_4 \eta^{\rho\sigma} \partial_\alpha h_{\lambda\beta} \partial^\lambda h^{\alpha\beta}
+ c_5 \eta^{\rho\sigma} \partial_\alpha h^{\alpha\beta} \partial_\beta h
\\
\\
+ d_1 h^{\rho\sigma} \partial_\alpha \partial^\alpha h
+ d_2 h^{\rho\sigma} \partial_\alpha \partial_\beta h^{\alpha\beta}
+ d_3 h \partial_\alpha \partial^\alpha h^{\rho\sigma}
+ d_4 h^{\alpha\beta} \partial_\alpha \partial_\beta h^{\rho\sigma}
+ d_{5_{i}} h^{\rho\alpha} \partial^\beta \partial_\beta h^{\sigma}_{\ \alpha} 
+ d_{5_{ii}} h^{\sigma\alpha} \partial^\beta \partial_\beta h^{\rho}_{\ \alpha}
\\
+ d_{6_{i}} h^{\rho\alpha} \partial_\alpha \partial_\beta h^{\sigma\beta} 
+ d_{6_{ii}} h^{\sigma\alpha} \partial_\alpha \partial_\beta h^{\rho\beta}
+ d_7 h \partial^\rho \partial^\sigma h
+ d_8 h_{\alpha\beta} \partial^\rho \partial^\sigma h^{\alpha\beta}
\\
+ d_{9_{i}} h^{\rho\alpha} \partial^\sigma \partial_\alpha h 
+ d_{9_{ii}} h^{\sigma\alpha} \partial^\rho \partial_\alpha h
+ d_{10_{i}} h^{\rho\alpha} \partial^\sigma \partial^\beta h_{\alpha\beta} 
+ d_{10_{ii}} h^{\sigma\alpha} \partial^\rho \partial^\beta h_{\alpha\beta}
\\
+ d_{11_{i}} h \partial^\rho \partial_\alpha h^{\sigma\alpha} 
+ d_{11_{ii}} h \partial^\sigma \partial_\alpha h^{\rho\alpha}
+ d_{12_{i}} h_{\alpha\beta} \partial^\rho \partial^\alpha h^{\sigma\beta} 
+ d_{12_{ii}} h_{\alpha\beta} \partial^\sigma \partial^\alpha h^{\rho\beta}
\\
+ a_1 \eta^{\rho\sigma} h_{\alpha\beta} \partial^\alpha \partial^\beta h
+ a_2 \eta^{\rho\sigma} h \partial_\alpha \partial_\beta h^{\alpha\beta}
+ a_3 \eta^{\rho\sigma} h_{\alpha\beta} \partial^\alpha \partial_\lambda h^{\lambda\beta}
+ a_4 \eta^{\rho\sigma} h \partial_\alpha \partial^\alpha h
+ a_5 \eta^{\rho\sigma} h_{\alpha\beta} \partial_\lambda \partial^\lambda h^{\alpha\beta}  , \label{genT}
\end{multline}

\noindent where we separate terms $h \partial \partial h$ that are proportional to the Minkowski metric $\eta^{\rho\sigma}$ with coefficients $a_n$. The general idea of the Fock method is to take the divergence $\partial_\rho T^{\rho\sigma}$ and solve for the free coefficients in front of each term such that the resulting energy-momentum tensor is conserved on-shell. These coefficients can be solved to impose various other properties, such as symmetry or tracelessness. Terms which must have an identical coefficient for a symmetric expression with subscripts (i) and (ii). For example, terms $b_{8_{i}}$ and $b_{8_{ii}}$ form a symmetric pair when $b_{8_{i}} = b_{8_{ii}}$. Terms $b_n$ correspond to terms $\partial h \partial h$ that are not proportional to Minkowski $\eta^{\rho\sigma}$, and terms $d_n$ correspond to terms $h \partial \partial h$ that are not proportional to Minkowski $\eta^{\rho\sigma}$. The four sets of free coefficients make the proofs and linear systems of equations easier to follow. \\

The general idea of the Fock method, to take the divergence $\partial_\rho T^{\rho\sigma}$ of (\ref{genT}) and solve for coefficients that allow for a conserved expression up to $E^{\mu\nu}$ in (\ref{FPEOM}). This will also include terms proportional to the trace of the equation of motion (\ref{FPEOM}) which we obtain from ${\bf{E}} = \eta_{\mu\nu} E^{\mu\nu}$,

\begin{equation}
{\bf{E}} =  \partial^\alpha   \partial^\beta h_{\alpha\beta} -  \square h  . \label{FPEOMtrace}
\end{equation}

We wish to explore the most general expression (\ref{genT}), and its relationship to the canonical Noether energy-momentum tensor (\ref{FPcanon}), supplemented by the most general possible superpotential and terms proportional to the equations of motion. There are six possible terms proportional to the equations of motion $h E^{\rho\sigma}$, $h^{\rho}_\alpha E^{\sigma\alpha}$, $h^{\rho\sigma} {\bf{E}}$, $h^{\sigma}_{\alpha} E^{\rho\alpha}$, $\eta^{\rho\sigma} h {\bf{E}}$ and $\eta^{\rho\sigma} h_{\alpha\beta} E^{\alpha\beta}$, each of which will be general up to a coefficient $\zeta_n$. Therefore what we will solve for is the most general possible case where one supplements the canonical Noether energy-momentum tensor of spin-2 Fierz-Pauli theory (\ref{FPcanon}) by the divergence of a superpotential and terms proportional to the equations of motion (i.e. we will obtain solutions of the form $T^{\rho\sigma} = T^{\rho\sigma}_C + \partial_\alpha \Psi^{[\rho\alpha]\sigma}
+ \zeta_1 h E^{\rho\sigma}
+ \zeta_2 h^{\rho}_\alpha E^{\sigma\alpha}
+ \zeta_3 h^{\rho\sigma} {\bf{E}}
+ \zeta_4 h^{\sigma}_{\alpha} E^{\rho\alpha}
+ \zeta_5 \eta^{\rho\sigma} h {\bf{E}}
+ \zeta_6 \eta^{\rho\sigma} h_{\alpha\beta} E^{\alpha\beta} $). To do this, we must re-express (\ref{genT}) in terms of the 28 possible terms from the $\zeta_n$ expressions, and all possible superpotential terms. This process is non-trivial, so we will now explain how one must re-express these terms.

In total there are 43 terms in (\ref{genT}). Using the identity $A\partial B=\partial (AB)-B \partial A$ based on the product rule for each of the quadratic terms $b_n$ and $c_n$, we can express all terms of the form $\partial h \partial h$ as $h \partial \partial h$ (terms presented in the equations of motion) plus terms under a total divergence of the form $\partial [h \partial h]$ (which contribute to the superpotential). For example we can re-write the $b_1$ term in (\ref{genT}) as $b_1 \partial_\alpha h^{\rho\sigma} \partial_\beta h^{\alpha\beta} = b_1 \partial_\alpha [h^{\rho\sigma} \partial_\beta h^{\alpha\beta}] - b_1 h^{\rho\sigma} \partial_\alpha \partial_\beta h^{\alpha\beta}$. No terms can be neglected as in the case of boundary terms in the action when deriving the equations of motion; total divergences in this derivation contribute to the superpotential. We give the result of this process in (\ref{genTsuper}), below. Terms of the form $h \partial \partial h$ will factor into combinations of the equations of motion in (\ref{FPEOM}) and (\ref{FPEOMtrace}). The total divergence term will result in the divergence of the superpotential term of the form $\partial_\alpha \Psi^{[\rho\alpha]\sigma}$. As discussed earlier, the superpotential $\Psi^{[\rho\alpha]\sigma}$ must be antisymmetric in $[\rho\alpha]$ so that the total expression for $T^{\rho\sigma}$ is conserved on-shell via $\partial_\rho \partial_\alpha \Psi^{[\rho\alpha]\sigma} = 0$ (since $T^{\rho\sigma}_C$ is independently conserved on-shell due to Noether's first theorem, and all of the terms proportional to the equations of motion trivially do not impact on-shell conservation).

The complicated part is that some $b_n$ and $c_n$ terms can be combined using the identity $A\partial B=\partial (AB)-B \partial A$ in two different ways. In addition there are terms $a_n$ and $d_n$ that the identity $A\partial B=\partial (AB)-B \partial A$ can be applied to twice, contributing two pieces to the superpotential and a different piece to the equations of motion. All of these possibilities must be accounted for in the most general linear system: one cannot simply split these possibilities by a numerical coefficient such as $\frac{1}{2}$, because the relative contribution can be uneven, such as in the case of the $b_2$ term in $T^{\mu\nu}_{LL}$ (\ref{LLEMT}), $- \frac{3}{2}  \partial_\alpha h^{\mu\nu}  \partial^\alpha h $. In general, we will split such terms in the form $b_n = B_n + \bar{B}_n $. For example term $b_2 = B_2 + \bar{B}_2$ must be split. This is because it can make contributions $B_2 \partial_\alpha h^{\rho\sigma} \partial^\alpha h = B_2 \partial_\alpha [h^{\rho\sigma} \partial^\alpha h] - B_2  h^{\rho\sigma} \partial_\alpha\partial^\alpha h$ and contribution $\bar{B}_2 \partial_\alpha h^{\rho\sigma} \partial^\alpha h = \bar{B}_2 \partial_\alpha  [ h \partial^\alpha h^{\rho\sigma}] - \bar{B}_2 h \partial_\alpha \partial^\alpha h^{\rho\sigma} $. For this reason terms $b_2$ can contribute multiple different terms to the superpotential of a particular energy-momentum tensor, as seen in the general result (\ref{genTsuper}), and as emphasized by the Landau-Lifshitz example. The exact splitting of each coefficient can be nontrivial and must be solved for as part of the general system of linear equations (which we summarize in Appendix A).

However, the majority of the terms in (\ref{genT}) cannot be split, because they will either not contribute to one of the 6 possible equations of motion (e.g. terms such as $h_{\alpha\beta} \partial^\alpha \partial^\beta h^{\rho\sigma}$), or they will produce a term symmetric in $(\rho\alpha)$ in the total divergence which cannot be incorporated into the superpotential $\Psi^{[\rho\alpha]\sigma}$ (e.g. terms with a $\partial^\rho$ total derivative). The third possibility is both applications of the identity $A\partial B=\partial (AB)-B \partial A$ yield the same result, thus they recombine and no splitting is necessary. Taking this all into account, there are 9 terms which must be split due to multiple possible contributions in the most general system. They are $a_1$, $a_2$, $b_2$, $b_4$, $c_5$, $d_1$, $d_3$, $d_{5_{i}}$ and $d_{5_{ii}}$. Each of these 9 coefficients is split in the form $a_n = A_n + \bar{A}_n $, $b_n = B_n + \bar{B}_n $ , $c_n = C_n + \bar{C}_n $ and $d_n = D_n + \bar{D}_n $. Using these conditions on (\ref{genT}) and the identity $A\partial B=\partial (AB)-B \partial A$ accordingly we are left with the general energy-momentum tensor,

\begin{multline}
T^{\rho\sigma} = (b_7 - d_8) \partial^\sigma h_{\alpha\beta} \partial^\rho h^{\alpha\beta}
+ c_4 \eta^{\rho\sigma} \partial_\alpha h_{\lambda\beta} \partial^\lambda h^{\alpha\beta}
+ (b_{10_{i}} - d_{12_{i}}) \partial_\alpha [\eta^{\rho\alpha}  h^{\sigma\beta} \partial^\omega h_{\omega\beta}  ]
\\
\\
+ \partial_\alpha [(B_2 + \bar{D}_1 - \bar{D}_3) h^{\rho\sigma} \partial^\alpha h
+  b_{9_{i}} h^{\sigma\alpha}  \partial^\rho h
+ (\bar{B}_2 - \bar{D}_1 + \bar{D}_3) h \partial^\alpha h^{\rho\sigma}
+ b_{8_{i}} h \partial^\rho h^{\sigma\alpha}
\\
+ (b_1 - d_4) h^{\rho\sigma} \partial_\beta h^{\alpha\beta}
+   b_3 h^{\sigma\alpha} \partial_\beta h^{\rho\beta}
+ d_4 h^{\alpha\beta} \partial_\beta h^{\rho\sigma}
+  b_5 h^{\rho\beta} \partial_\beta h^{\sigma\alpha}
\\
+  d_{12_{i}} h_{\ \beta}^{\alpha}  \partial^\rho h^{\sigma\beta}
+ (B_4 +  \bar{D}_{5_{i}} - \bar{D}_{5_{ii}}) h^{\rho}_{\ \beta} \partial^\alpha h^{\sigma\beta}
+ d_{12_{ii}}  h_{\ \beta}^{\alpha} \partial^\sigma h^{\rho\beta}
+ b_{11_{ii}}  h^{\rho}_{\ \beta} \partial^\sigma h^{\alpha\beta}
\\
+  (\bar{B}_4 -  \bar{D}_{5_{i}} + \bar{D}_{5_{ii}}) h^{\sigma}_{\ \beta} \partial^\alpha h^{\rho\beta}
+  b_{11_{i}} h^{\sigma}_{\ \beta}  \partial^\rho h^{\alpha\beta}
+ b_6  \eta^{\sigma\alpha} h \partial^\rho h
+ c_1  \eta^{\rho\sigma} h \partial^\alpha h
\\
+ d_8 \eta^{\sigma\alpha} h^{\omega\beta} \partial^\rho h_{\omega\beta}
+ c_2 \eta^{\rho\sigma} h_{\beta\lambda} \partial^\alpha h^{\beta\lambda}
+  b_{8_{ii}} \eta^{\sigma\alpha} h^{\rho\omega} \partial_\omega h
+ (C_5 + \bar{A}_1 - \bar{A}_2) \eta^{\rho\sigma} h^{\alpha\beta} \partial_\beta h
\\
+  (\bar{C}_5 - \bar{A}_1 + \bar{A}_2) \eta^{\rho\sigma} h \partial_\omega h^{\omega\alpha}
+    b_{9_{ii}} \eta^{\sigma\alpha} h \partial_\omega h^{\rho\omega}
+ (b_{10_{ii}}  - d_{12_{ii}}) \eta^{\sigma\alpha} h^{\rho\omega} \partial^\beta h_{\omega\beta}
+  c_3 \eta^{\rho\sigma} h^{\alpha\beta} \partial_\lambda h^{\lambda}_{\ \beta}
]
\\
\\
+ h [(d_7 - b_6) \partial^\rho \partial^\sigma h
+ (D_3 - \bar{B}_2 + \bar{D}_1) \partial_\alpha \partial^\alpha h^{\rho\sigma}
+ (d_{11_{i}} -  b_{8_{i}} )  \partial^\rho \partial_\alpha h^{\sigma\alpha} 
\\
+ (d_{11_{ii}} - b_{9_{ii}} )  \partial^\sigma \partial_\alpha h^{\rho\alpha}
+ \bar{M}_3 \eta^{\rho\sigma}  \partial_\alpha \partial_\beta h^{\alpha\beta}
+ \bar{M}_4 \eta^{\rho\sigma}   \partial_\alpha \partial^\alpha h ]
\\
\\
+ h^{\rho}_{\alpha} [(d_{10_{i}} + d_{12_{ii}} - b_{10_{ii}} - b_{11_{ii}} ) \partial^\sigma \partial_\beta h^{\alpha\beta} 
+ (d_{9_{i}}  -  b_{8_{ii}} )  \partial^\sigma \partial^\alpha h 
+ (d_{6_{i}}  -  b_5)  \partial^\alpha \partial_\beta h^{\sigma\beta}
\\
+ (D_{5_{i}} + \bar{D}_{5_{ii}} - B_4)  \partial^\beta \partial_\beta h^{\sigma\alpha} 
+ \bar{M}_1 \eta^{\sigma\alpha} \partial_\omega \partial_\beta h^{\omega\beta}
+ \bar{M}_2 \eta^{\sigma\alpha} \partial_\omega \partial^\omega h]
\\
\\
+ h^{\rho\sigma} [M_1 \partial_\alpha \partial_\beta h^{\alpha\beta}
+ M_2 \partial_\alpha \partial^\alpha h
]
\\
\\
+ \eta^{\rho\sigma} h [M_3 \partial_\alpha \partial_\beta h^{\alpha\beta}
+ M_4 \partial_\alpha \partial^\alpha h
]
\\
\\
+ h^{\sigma}_{\alpha} [(d_{10_{ii}} + d_{12_{i}}  -  b_{10_{i}}  -  b_{11_{i}}) \partial^\rho \partial_\beta h^{\alpha\beta}
+ (d_{9_{ii}} -  b_{9_{i}})  \partial^\rho \partial^\alpha h
+ (d_{6_{ii}} -  b_3 ) \partial^\alpha \partial_\beta h^{\rho\beta}
\\
+ (D_{5_{ii}} + \bar{D}_{5_{i}} - \bar{B}_4) \partial^\beta \partial_\beta h^{\rho\alpha}
+ \hat{M}_1 \eta^{\rho\alpha}\partial_\omega \partial_\beta h^{\omega\beta}
+ \hat{M}_2 \eta^{\rho\alpha}\partial_\omega \partial^\omega h] 
\\
\\
+ \eta^{\rho\sigma} h_{\alpha\beta} [(a_5 - c_2 ) \partial_\lambda \partial^\lambda h^{\alpha\beta}
+ (A_1 - C_5 + \bar{A}_2)   \partial^\alpha \partial^\beta h
+ \frac{1}{2} (a_3 - c_3 )  \partial_\omega \partial^\alpha h^{\omega\beta}
\\
+ \frac{1}{2} (a_3 - c_3  )  \partial_\omega \partial^\beta h^{\omega\alpha}
+ \hat{M}_3 \eta^{\alpha\beta}  \partial_\omega \partial_\gamma h^{\omega\gamma}
+ \hat{M}_4  \eta^{\alpha\beta} \partial_\omega \partial^\omega h] .
\label{genTsuper}
\end{multline}

\noindent The $M_n$ coefficients are required because after separating terms, an additional splitting is required for terms proportional to the Minkowski metric $\eta^{\rho\sigma}$ that can each be separated in 3 possible ways across the equations of motion. They were separated according to:

\begin{gather}
d_2 + d_4 - b_1 = M_1 + \bar{M}_1 + \hat{M}_1 \label{M1gen}
\\
D_1 - B_2 + \bar{D}_3 = M_2 + \bar{M}_2 + \hat{M}_2 \label{M2gen}
\\
A_2 - \bar{C}_5 +  \bar{A}_1 = M_3 + \bar{M}_3 + \hat{M}_3 \label{M3gen}
\\
a_4 - c_1 = M_4 + \bar{M}_4 + \hat{M}_4 \label{M4gen}
\end{gather}

Solving the linear system of equations for the coefficients in (\ref{genTsuper}) and imposing the conditions on $T^{\rho\sigma}$ gives insight into the most general energy-momentum tensor for linearized gravity written in terms of the canonical Noether $T^{\rho\sigma}_C$ of spin-2 Fierz-Pauli theory plus the divergence of a superpotential and terms proportional to the equations of motion. The bottom 6 groups of terms in (\ref{genTsuper}) are the six possible terms proportional to the equations of motion. The total divergence on the second line of (\ref{genTsuper}) is sorted according to pairs which will form the most general possible superpotential $\Psi^{[\rho\alpha]\sigma}$ for linearized gravity according to the most general Fock expression in (\ref{genT}).

We note the 3 terms, separated at the top of the (\ref{genTsuper}) expression, cannot be fit into either the most general superpotential or terms proportional to the equations of motion. The first two terms $(b_7 - d_8) \partial^\sigma h_{\alpha\beta} \partial^\rho h^{\alpha\beta}$ and $c_4 \eta^{\rho\sigma} \partial_\alpha h_{\lambda\beta} \partial^\lambda h^{\alpha\beta}$ are found in the canonical energy-momentum tensor (\ref{FPcanon}) which, in part, explains why ad-hoc addition of the divergence of a superpotential and terms proportional to the equations of motion can seemingly be used to obtain any published energy-momentum tensor. The final term $(b_{10_{i}} - d_{12_{i}}) \partial_\alpha [\eta^{\rho\alpha}  h^{\sigma\beta} \partial^\omega h_{\omega\beta}  ]$ is symmetric in $(\rho\alpha)$ in the total divergence thus cannot be combined to the superpotential, and it cannot be combined into any of the equations of motion. This will produce an independent equation in our general linear system.

In addition the manifestly symmetric form follows from the symmetry conditions:

\begin{gather}
b_n = b_{n_{i}} = b_{n_{ii}}  \label{symm1}
\\
d_n = d_{n_{i}} = d_{n_{ii}}  \label{symm2}
\end{gather}

\noindent We will return to these symmetry conditions later in the article. 

\section{3) The most general canonical Noether energy-momentum tensor supplemented by ad-hoc ``improvements''}

We will now ask the general question, namely, what is the most general possible superpotential, and terms proportional to the equations of motion, that can be added ad-hoc to the canonical Noether energy-momentum tensor (\ref{FPcanon}) in order to obtain a general system of on-shell conserved energy-momentum tensors in linearized gravity. In order to not impact the flow of the text, we present the system of linear equations of the coefficients resulting from this process in Appendix A. Any energy-momentum tensor for linearized gravity that can be obtained by ``improving'' the canonical Noether tensor (\ref{FPcanon}) can be determined by solving this system of linear equations; the exact superpotential and terms proportional to the equations of motion (\ref{FPEOM}) trivially follow. 

The aforementioned first 3 terms in (\ref{genTsuper}) are independent conditions that must be solved. The first two, as appearing in (\ref{FPcanon}), must be related to the coefficients of the spin-2 Fierz-Pauli canonical energy-momentum tensor (which appears the same in both (\ref{FPhilbert}) and (\ref{LLEMT}) as well). The third term must be independently satisfied. Therefore to obtain the canonical $T^{\rho\sigma}_C$ (\ref{FPcanon}) in (\ref{genTsuper}) we minimally require the conditions in equations  (\ref{canoncondi1}) to (\ref{canoncondi3}) in Appendix A.

The remaining terms in (\ref{FPcanon}) must be extracted from the general system of coefficients. Since (\ref{FPcanon}) has coefficients from (\ref{genT}) that are $c_1 = \frac{1}{4}$, $c_2 = - \frac{1}{4}$, $c_4 = \frac{1}{2}$, $c_5 = -\frac{1}{2}$, $b_6 = - \frac{1}{2}$, $b_7 = \frac{1}{2}$, $b_{8_{ii}} = \frac{1}{2}$, $b_{9_{ii}} = \frac{1}{2}$, and $b_{11_{ii}} = -1$, and we have already solved for $c_4$ and $b_7$, we only need to extract the remaining coefficients. Thus we need to shift the coefficients in (\ref{genTsuper}) by $c_1 \to c_1-\frac{1}{4}$, $c_2 \to  c_2+\frac{1}{4}$,  $c_5 \to c_5+\frac{1}{2}$, $b_6 \to b_6+\frac{1}{2}$, $b_{8_{ii}} \to b_{8_{ii}}-\frac{1}{2}$, $b_{9_{ii}} \to b_{9_{ii}}- \frac{1}{2}$, and $b_{11_{ii}} \to b_{11_{ii}}+1$ to exactly obtain (\ref{FPcanon}) in (\ref{genTsuper}). This modifies the $c_5$ splitting condition, thus we now have the splitting conditions from (\ref{splitcanoncondi1}) to (\ref{splitcanoncondi9}).

These coefficient shifts (obtained by extracting the canonical Noether energy-momentum tensor) modify the general superpotential in (\ref{genTsuper}) to,

\begin{multline}
\Psi^{[\rho\alpha]\sigma} = (B_2 + \bar{D}_1 - \bar{D}_3) h^{\rho\sigma} \partial^\alpha h
+  b_{9_{i}} h^{\sigma\alpha}  \partial^\rho h
+ (\bar{B}_2 - \bar{D}_1 + \bar{D}_3) h \partial^\alpha h^{\rho\sigma}
+ b_{8_{i}} h \partial^\rho h^{\sigma\alpha}
\\
+ (b_1 - d_4) h^{\rho\sigma} \partial_\beta h^{\alpha\beta}
+   b_3 h^{\sigma\alpha} \partial_\beta h^{\rho\beta}
+ d_4 h^{\alpha\beta} \partial_\beta h^{\rho\sigma}
+  b_5 h^{\rho\beta} \partial_\beta h^{\sigma\alpha}
\\
+  d_{12_{i}} h_{\ \beta}^{\alpha}  \partial^\rho h^{\sigma\beta}
+ (B_4 +  \bar{D}_{5_{i}} - \bar{D}_{5_{ii}}) h^{\rho}_{\ \beta} \partial^\alpha h^{\sigma\beta}
+ d_{12_{ii}}  h_{\ \beta}^{\alpha} \partial^\sigma h^{\rho\beta}
+ (b_{11_{ii}}+1)  h^{\rho}_{\ \beta} \partial^\sigma h^{\alpha\beta}
\\
+  (\bar{B}_4 -  \bar{D}_{5_{i}} + \bar{D}_{5_{ii}}) h^{\sigma}_{\ \beta} \partial^\alpha h^{\rho\beta}
+  b_{11_{i}} h^{\sigma}_{\ \beta}  \partial^\rho h^{\alpha\beta}
+ (b_6+\frac{1}{2})  \eta^{\sigma\alpha} h \partial^\rho h
+ (c_1-\frac{1}{4})  \eta^{\rho\sigma} h \partial^\alpha h
\\
+ d_8 \eta^{\sigma\alpha} h^{\omega\beta} \partial^\rho h_{\omega\beta}
+ (c_2+\frac{1}{4}) \eta^{\rho\sigma} h_{\beta\lambda} \partial^\alpha h^{\beta\lambda}
+  (b_{8_{ii}}-\frac{1}{2}) \eta^{\sigma\alpha} h^{\rho\omega} \partial_\omega h
+ (C_5 + \bar{A}_1 - \bar{A}_2) \eta^{\rho\sigma} h^{\alpha\beta} \partial_\beta h
\\
+  (\bar{C}_5 - \bar{A}_1 + \bar{A}_2) \eta^{\rho\sigma} h \partial_\omega h^{\omega\alpha}
+   (b_{9_{ii}}- \frac{1}{2}) \eta^{\sigma\alpha} h \partial_\omega h^{\rho\omega}
+ (b_{10_{ii}}  - d_{12_{ii}}) \eta^{\sigma\alpha} h^{\rho\omega} \partial^\beta h_{\omega\beta}
+  c_3 \eta^{\rho\sigma} h^{\alpha\beta} \partial_\lambda h^{\lambda}_{\ \beta} .
\label{canonsupergen}
\end{multline}

\noindent We note that antisymmetric pairs are sorted throughout this expression. Using the superpotential condition $\partial_\rho \partial_\alpha \Psi^{[\rho\alpha]\sigma} = 0$ we therefore straightforwardly obtain the conditions for the antisymmetric superpotential in equations (\ref{canonsuper1}) to (\ref{canonsuper12}).

Imposing conditions (\ref{canoncondi1}) to (\ref{canoncondi3}), (\ref{splitcanoncondi1}) to (\ref{splitcanoncondi9}) and (\ref{canonsuper1}) to (\ref{canonsuper12}) from Appendix A on (\ref{genTsuper}) and writing the equation of motion coefficients in terms of $\zeta_n$ we obtain the desired compact result,

\begin{equation}
T^{\rho\sigma} = T^{\rho\sigma}_C + \partial_\alpha \Psi^{[\rho\alpha]\sigma}
+ \zeta_1 h E^{\rho\sigma}
+ \zeta_2 h^{\rho}_\alpha E^{\sigma\alpha}
+ \zeta_3 h^{\rho\sigma} {\bf{E}}
+ \zeta_4 h^{\sigma}_{\alpha} E^{\rho\alpha}
+ \zeta_5 \eta^{\rho\sigma} h {\bf{E}}
+ \zeta_6 \eta^{\rho\sigma} h_{\alpha\beta} E^{\alpha\beta} ,
\label{genadhoc}
\end{equation}

\noindent where the equations of motion have slightly modified coefficients when compared to those in (\ref{genTsuper}) due to the coefficient shifts above. The resulting system of linear equations for the equations of motion $\zeta_n$ are given in Appendix A as (\ref{canoneomcondi1}) to (\ref{canoneomcondi28}). The $M$ coefficients in (\ref{M1gen}) to (\ref{M4gen}) have also been modified from the canonical Noether coefficients, given in Appendix A as (\ref{Mcanon1}) to (\ref{Mcanon4}).

Therefore we now have the most general ``improvement'' of the canonical Noether tensor in (\ref{genadhoc}), with all possible superpotentials (\ref{canonsupergen}) and all possible terms proportional to the equations of motion that can be added. These were directly derived from (\ref{genT}), therefore we have a direct connection between any linearized gravity energy-momentum tensor, and all which can be obtained by ad-hoc improving the canonical Noether expression in (\ref{FPcanon}). By solving the system of linear equations in Appendix A, one finds solutions which satisfy both criteria, that the conserved energy-momentum tensor in (\ref{genT}) will be derivable from (\ref{FPcanon}) supplemented by the divergence of a superpotential and terms proportional to the equations of motion in (\ref{FPEOM}) and (\ref{FPEOMtrace}), as given in (\ref{genadhoc}). This leads us to our first result: there are infinitely many solutions to the linear system in Appendix A. In other words, there are infinitely many divergences of superpotentials and terms proportional to the equations of motion that can be added to the canonical Noether energy-momentum tensor (\ref{FPcanon}) in order to obtain on-shell conserved tensors for linearized gravity. If in addition we use the symmetry conditions in (\ref{symm1}) and (\ref{symm2}) we find that there are infinitely many $T^{\rho\sigma}$ which are both symmetric and conserved. The ``improvements'' used to obtain e.g. the Hilbert and Landau-Liftshitz energy-momentum tensors in (\ref{converseresult}) and (\ref{LLadhocimproved}) from $T^{\rho\sigma}_C$ are not special or unique; they are just two of infinitely many possibly solutions. This suggests claims of a meaningful connection of a given $T^{\rho\sigma}$ to Noether's first theorem using the ad-hoc ``improvement'' method should not be made.

To recap, we will summarize the equations in Appendix A that give the conditions necessary for linearized gravity energy-momentum tensors of the form (\ref{genadhoc}). Equations (\ref{canoncondi1}) to (\ref{canoncondi3}) give the conditions necessary for the most general linearized gravity energy-momentum tensor (\ref{genT}) to be expressed as the canonical Noether energy-momentum tensor improved by the divergence of a superpotential and terms proportional to the equations of motion. Equations (\ref{splitcanoncondi1}) to (\ref{splitcanoncondi9}) are the conditions for coefficient splitting modified by the canonical Noether tensor. Equations (\ref{canonsuper1}) to (\ref{canonsuper12}) are the conditions required to have a superpotential antisymmetric in $[\rho\alpha]$. Equations (\ref{canoneomcondi1}) to (\ref{canoneomcondi28}) are the conditions for each of the 6 $\zeta_n$ equations of motion, and (\ref{Mcanon1}) to (\ref{Mcanon4}) are the conditions on the $M_n$ terms within. Finally if we wish to derive symmetric expressions we can use the symmetry conditions (\ref{symm1}) and (\ref{symm2}) from earlier in the article.

\section{4) Verifying the general system for the Hilbert and Landau-Lifshitz energy-momentum tensors}

We will now use our motivating examples (Hilbert and Landau-Lifshitz energy-momentum tensors) to apply the general results.

\subsection{4.1) Hilbert coefficients and solution}

For the Hilbert coefficients in (\ref{FPhilbert}), from (\ref{genT}) we have $c_1 = \frac{1}{4}$, $c_2 = - \frac{1}{4}$, $c_4 = \frac{1}{2}$, $a_1 = \frac{1}{2}$, $b_{11_{i}} = -1$, $b_{11_{ii}} = -1$, $b_4 = 1$, $b_5 = 1$, $b_1 = -1$, $b_2 = -\frac{1}{2}$, $b_6 = -\frac{1}{2}$, $b_7 = \frac{1}{2}$, $b_{9_{ii}} = \frac{1}{2}$, $b_{9_{i}} = \frac{1}{2}$, $d_{6_{i}} = 1$, $d_{6_{ii}} = 1$, $d_2 = -1$, $d_4 = -1$, $d_1 = \frac{1}{2}$, $d_{9_{i}} = -\frac{1}{2}$, and $d_{9_{ii}} = -\frac{1}{2}$. These satisfy the symmetry conditions (\ref{symm1}) and (\ref{symm2}).

This is a solution to the linear system in Appendix A, with $\zeta_4 = -2$, and all other $\zeta_n = 0$. These coefficients fix the general superpotential in (\ref{canonsupergen}) to yield $\Psi^{[\rho\alpha]\sigma} = \frac{1}{2} h^{\rho\sigma} \partial^\alpha h
- \frac{1}{2} h^{\sigma\alpha}  \partial^\rho h
-  h^{\alpha\beta} \partial_\beta h^{\rho\sigma}
+   h^{\rho\beta} \partial_\beta h^{\sigma\alpha}
+ h^{\sigma}_{\ \beta} \partial^\alpha h^{\rho\beta}
- h^{\sigma}_{\ \beta}  \partial^\rho h^{\alpha\beta}
-\frac{1}{2} \eta^{\sigma\alpha} h^{\rho\omega} \partial_\omega h
+ \frac{1}{2} \eta^{\rho\sigma} h^{\alpha\beta} \partial_\beta h$. This is exactly the well known Hilbert superpotential in (\ref{conversesuperpotential}). Substituting these solutions back into (\ref{genadhoc}) we immediately obtain the well-known result for the Hilbert energy-momentum tensor (\ref{converseresult}). Therefore the general system of equations in Appendix A recovers the Hilbert result.

\subsection{4.2) Landau-Lifshitz coefficients and solution}

For the Landau-Lifshitz coefficients in (\ref{LLEMT}), from (\ref{genT}) we have $c_1 = \frac{3}{4}$, 
$c_2 = - \frac{1}{4}$, 
$c_4 = \frac{1}{2}$, 
$c_5 = - 1$, 
$b_6 = - 1$, 
$b_2 = -\frac{3}{2}$, 
$b_1 = 1$, 
$b_3 = -1$, 
$b_7 = \frac{1}{2}$, 
$b_4 = 1$, 
$b_{9_{ii}} = 1$, 
$b_{9_{i}} = 1$, 
$b_{8_{i}} = \frac{1}{2}$, 
$b_{8_{ii}} = \frac{1}{2}$, 
$b_{11_{i}} = -1$, 
and $b_{11_{ii}} = -1$. These satisfy the symmetry conditions (\ref{symm1}) and (\ref{symm2}).

This is a solution to the linear system in Appendix A, with $\zeta_1 = 1$, $\zeta_4 = -2$, and all other $\zeta_n = 0$. These coefficients fix the general superpotential in (\ref{canonsupergen}) to yield $\Psi^{[\rho\alpha]\sigma} = - h^{\rho\sigma} \partial^\alpha h
+ h^{\sigma\alpha}  \partial^\rho h
- \frac{1}{2} h \partial^\alpha h^{\rho\sigma}
+ \frac{1}{2} h \partial^\rho h^{\sigma\alpha}
+ h^{\rho\sigma} \partial_\beta h^{\alpha\beta}
- h^{\sigma\alpha} \partial_\beta h^{\rho\beta}
+  h^{\sigma}_{\ \beta} \partial^\alpha h^{\rho\beta}
- h^{\sigma}_{\ \beta}  \partial^\rho h^{\alpha\beta}
- \frac{1}{2}  \eta^{\sigma\alpha} h \partial^\rho h
+ \frac{1}{2} \eta^{\rho\sigma} h \partial^\alpha h
- \frac{1}{2} \eta^{\rho\sigma} h \partial_\omega h^{\omega\alpha}
+  \frac{1}{2} \eta^{\sigma\alpha} h \partial_\omega h^{\rho\omega}$. This is exactly the well known Landau-Lifshitz superpotential (\ref{LLknownsuper}). Substituting these solutions back into (\ref{genadhoc}) we immediately obtain the well known result for the Landau-Lifshitz energy-momentum tensor in (\ref{LLadhocimproved}). Therefore the general system of equations in Appendix A recovers the Landau-Lifshitz result.

\section{5) Two new energy-momentum tensors derivable from ad-hoc improving the canonical Noether energy-momentum tensor}

We now derive two new energy-momentum tensors\footnote{We will call the new expressions the Audrey and Elizabeth energy-momentum tensors, named after our Grandmothers.}  that can be obtained from improving the canonical Noether tensor by solving the system of equations in Appendix A, just like the Hilbert (\ref{FPhilbert}) and Landau-Lifshitz (\ref{LLEMT}) expressions.

\subsection{5.1) Elizabeth energy-momentum tensor}

For the Elizabeth energy-momentum tensor $T^{\rho\sigma}_E$, we will use the symmetry conditions (\ref{symm1}) and (\ref{symm2}). We find a symmetric solution to the linear system in Appendix A to be $c_1 =  \frac{1}{4}$, $c_2 = - \frac{1}{4}$, $c_4 = \frac{1}{2}$,
$b_1 = 1$, $b_2 = -\frac{1}{2}$, $b_3 = -1$, $b_4 = 1$, $b_6 = - \frac{1}{2}$, $b_7 = \frac{1}{2}$,  $b_9 = \frac{1}{2}$, $b_{11} = -1$,  
$a_1 = \frac{1}{2}$, $a_2 = -\frac{1}{2}$, $a_4 = \frac{1}{2}$, 
$d_1 = \frac{1}{2}$, $d_9 = - \frac{1}{2}$. This yields from (\ref{genT}) the energy-momentum tensor,

\begin{multline}
T^{\rho\sigma}_E = 
 \partial_\alpha h^{\rho\sigma} \partial_\beta h^{\alpha\beta}
- \frac{1}{2} \partial_\alpha h^{\rho\sigma} \partial^\alpha h 
- \partial_\alpha h^{\rho\alpha} \partial_\beta h^{\sigma\beta}
+ \partial_\alpha h^{\rho}_{\ \beta} \partial^\alpha h^{\sigma\beta}
\\
- \frac{1}{2} \partial^\rho h \partial^\sigma h 
+ \frac{1}{2} \partial^\rho h_{\alpha\beta} \partial^\sigma h^{\alpha\beta}
+ \frac{1}{2} \partial^\rho h \partial_\alpha h^{\sigma\alpha} 
+ \frac{1}{2} \partial^\sigma h \partial_\alpha h^{\rho\alpha}
- \partial^\rho h_{\alpha\beta} \partial^\alpha h^{\sigma\beta} 
- \partial^\sigma h_{\alpha\beta} \partial^\alpha h^{\rho\beta}
\\
+ \frac{1}{4} \eta^{\rho\sigma} \partial_\alpha h \partial^\alpha h
- \frac{1}{4} \eta^{\rho\sigma} \partial_\alpha h_{\beta\lambda} \partial^\alpha h^{\beta\lambda}
+ \frac{1}{2} \eta^{\rho\sigma} \partial_\alpha h_{\lambda\beta} \partial^\lambda h^{\alpha\beta}
\\
+ \frac{1}{2} h^{\rho\sigma} \partial_\alpha \partial^\alpha h
- \frac{1}{2} h^{\rho\alpha} \partial^\sigma \partial_\alpha h 
- \frac{1}{2} h^{\sigma\alpha} \partial^\rho \partial_\alpha h
+ \frac{1}{2} \eta^{\rho\sigma} h_{\alpha\beta} \partial^\alpha \partial^\beta h
- \frac{1}{2} \eta^{\rho\sigma} h \partial_\alpha \partial_\beta h^{\alpha\beta}
+ \frac{1}{2} \eta^{\rho\sigma} h \partial_\alpha \partial^\alpha h . \label{ElizabethT}
\end{multline}

\noindent This fixes $\zeta_4 = - 2$ and $\zeta_5 = - \frac{1}{2}$ in (\ref{genadhoc}), with the remaining $\zeta_n = 0$. From (\ref{canonsupergen}) we obtain the superpotential,

\begin{multline}
\Psi^{[\rho\alpha]\sigma}_E = - \frac{1}{2} h^{\rho\sigma} \partial^\alpha h
+  \frac{1}{2} h^{\sigma\alpha}  \partial^\rho h
+  h^{\rho\sigma} \partial_\beta h^{\alpha\beta}
- h^{\sigma\alpha} \partial_\beta h^{\rho\beta}
\\
+  h^{\sigma}_{\ \beta} \partial^\alpha h^{\rho\beta}
-  h^{\sigma}_{\ \beta}  \partial^\rho h^{\alpha\beta}
-\frac{1}{2} \eta^{\sigma\alpha} h^{\rho\omega} \partial_\omega h
+ \frac{1}{2} \eta^{\rho\sigma} h^{\alpha\beta} \partial_\beta h .
\end{multline}

\noindent Thus (\ref{ElizabethT}) can be derived from the canonical Noether energy-momentum tensor (\ref{FPcanon}) by adding ad-hoc $\partial_\alpha \Psi^{[\rho\alpha]\sigma}_E$ and $- 2 h^{\sigma}_{\alpha} E^{\rho\alpha} - \frac{1}{2}  \eta^{\rho\sigma} h {\bf{E}}$,

\begin{equation}
T^{\rho\sigma}_E = T^{\rho\sigma}_C
+ \partial_\alpha \Psi^{[\rho\alpha]\sigma}_E
- 2 h^{\sigma}_{\alpha} E^{\rho\alpha}
- \frac{1}{2}  \eta^{\rho\sigma} h {\bf{E}} .
\end{equation}

Similarly, we can obtain infinitely many conserved, symmetric energy-momentum tensors for linearized gravity from the canonical Noether energy-momentum tensor (two of which are the Hilbert (\ref{FPhilbert}) and Landau-Lifshitz (\ref{LLEMT}) expressions).

\subsection{5.2) Audrey energy-momentum tensor}

For the Audrey energy-momentum tensor $T^{\rho\sigma}_A$, we will use the conditions $a_n = 0$ and $b_n = 0$. Using these conditions we will prove that no symmetric expressions exist without these terms. However, a conserved expression can be derived by improving the canonical Noether energy-momentum tensor (\ref{FPcanon}). We find a solution to the coefficients in Appendix A with $a_n = 0$ and $b_n = 0$ to be $c_1 = - \frac{1}{4}$, $c_2 =  \frac{1}{4}$, $c_3 = - 1$, $c_4 = \frac{1}{2}$, $c_5 = \frac{1}{2}$, $d_8 = - \frac{1}{2}$. $d_{12_{ii}} = -1$,  $d_7 = \frac{1}{2}$, $d_{11_{ii}} = -\frac{1}{2}$, $d_{9_{i}} = -\frac{1}{2}$ and $d_{10_{i}} = 2$. This yields from (\ref{genT}) the energy-momentum tensor,

\begin{multline}
T^{\rho\sigma}_A = 
- \frac{1}{4} \eta^{\rho\sigma} \partial_\alpha h \partial^\alpha h
+ \frac{1}{4} \eta^{\rho\sigma} \partial_\alpha h_{\beta\lambda} \partial^\alpha h^{\beta\lambda}
- \eta^{\rho\sigma} \partial_\alpha h^{\alpha\beta} \partial_\lambda h^{\lambda}_{\ \beta}
+ \frac{1}{2} \eta^{\rho\sigma} \partial_\alpha h_{\lambda\beta} \partial^\lambda h^{\alpha\beta}
+ \frac{1}{2} \eta^{\rho\sigma} \partial_\alpha h^{\alpha\beta} \partial_\beta h
\\
+ \frac{1}{2} h \partial^\rho \partial^\sigma h
- \frac{1}{2} h_{\alpha\beta} \partial^\rho \partial^\sigma h^{\alpha\beta}
- \frac{1}{2} h^{\rho\alpha} \partial^\sigma \partial_\alpha h 
- \frac{1}{2} h \partial^\sigma \partial_\alpha h^{\rho\alpha}
- h_{\alpha\beta} \partial^\sigma \partial^\alpha h^{\rho\beta}
+ 2 h^{\rho\alpha} \partial^\sigma \partial^\beta h_{\alpha\beta} . 
\label{AudreyT}
\end{multline}

\noindent This fixes $\zeta_6 = - 1$ in (\ref{genadhoc}), with the remaining $\zeta_n = 0$. From (\ref{canonsupergen}) we obtain the superpotential,

\begin{multline}
\Psi^{[\rho\alpha]\sigma}_A = 
-  h_{\ \beta}^{\alpha} \partial^\sigma h^{\rho\beta}
+   h^{\rho}_{\ \beta} \partial^\sigma h^{\alpha\beta}
+ \frac{1}{2}  \eta^{\sigma\alpha} h \partial^\rho h
- \frac{1}{2}  \eta^{\rho\sigma} h \partial^\alpha h
- \frac{1}{2} \eta^{\sigma\alpha} h^{\omega\beta} \partial^\rho h_{\omega\beta}
+ \frac{1}{2} \eta^{\rho\sigma} h_{\beta\lambda} \partial^\alpha h^{\beta\lambda}
\\
 -\frac{1}{2} \eta^{\sigma\alpha} h^{\rho\omega} \partial_\omega h
+ \frac{1}{2} \eta^{\rho\sigma} h^{\alpha\beta} \partial_\beta h
+  \frac{1}{2} \eta^{\rho\sigma} h \partial_\omega h^{\omega\alpha}
- \frac{1}{2} \eta^{\sigma\alpha} h \partial_\omega h^{\rho\omega}
+  \eta^{\sigma\alpha} h^{\rho\omega} \partial^\beta h_{\omega\beta}
- \eta^{\rho\sigma} h^{\alpha\beta} \partial_\lambda h^{\lambda}_{\ \beta} .
\end{multline}

\noindent Thus (\ref{AudreyT}) can be derived from the canonical Noether energy-momentum tensor (\ref{FPcanon}) by adding ad-hoc $\partial_\alpha \Psi^{[\rho\alpha]\sigma}_A$ and $- \eta^{\rho\sigma} h_{\alpha\beta} E^{\alpha\beta}$,

\begin{equation}
T^{\rho\sigma}_A = T^{\rho\sigma}_C
+ \partial_\alpha \Psi^{[\rho\alpha]\sigma}_A
- \eta^{\rho\sigma} h_{\alpha\beta} E^{\alpha\beta} .
\end{equation}

We cannot have a symmetric expression here because $a_n = 0$ and $b_n = 0$ fixes $d_{12_{i}} = 0$ and $d_{12_{ii}} = -1$ which breaks the symmetry conditions (\ref{symm1}) and (\ref{symm2}). Both conditions must hold in order to have a symmetric energy-momentum tensor in (\ref{genT}).

\section{6) There exist infinitely many symmetric, conserved energy-momentum tensors from the Belinfante superpotential alone}

We now present, perhaps our most significant result, that the Belinfante superpotential is associated with infinitely many symmetric and conserved linearized gravity energy-momentum tensors. This is an important result because despite the various possible superpotentials one can add (such as $\Psi^{[\rho\alpha]\sigma}_{LL}$ or those found in \cite{bicak2016}), the Belinfante superpotential is the most commonly published expression. Our result is contrary to popular belief in the recent literature that the Hilbert energy-momentum tensor uniquely specifies the Belinfante energy-momentum tensor \cite{deser2010,barcelo2014}. This point is central to the recent Padmanabhan-Deser debate \cite{padmanabhan2008,deser2010}, in which the authors have argued about whether or not general relativity can be derived from spin-2 using a $T^{\mu\nu}$ resulting from Noether's theorem. Deser claimed that ad-hoc improving the canonical Noether tensor (\ref{FPcanon}) with the Belinfante superpotential uniquely gives the Hilbert energy-momentum tensor (\ref{FPhilbert}), thus he argued one does not have to use Noether's theorem at all to have a result from Noether's theorem, they can simply use the Hilbert approach. Such claims of general equivalence of the Noether and Hilbert methods for deriving an energy-momentum tensor has since been disproved in \cite{baker2021}. Deser's assertions come from results that have a long history \cite{rosenfeld1940,gotay1992,babak1999,borokhov2002,saravi2004,forger2004,leclerc2006b,leclerc2006,pons2018} of investigating the relationship between the Belinfante and Hilbert energy-momentum tensors. The general conclusion in the literature is that one can add the divergence of the Belinfante superpotential and terms proportional to the equation of motion to reconcile the Belinfante and Hilbert definitions (this was confirmed by our results). But this does not prove uniqueness! Indeed the Belinfante superpotential coincides with what we found for Hilbert in (\ref{conversesuperpotential}). However, as we will prove, this result is not unique because there are infinitely many symmetric and conserved expressions associated to this particular superpotential alone; infinitely many combinations of the equations of motion in (\ref{genadhoc}) are solutions to the system of equations in Appendix A when the Belinfante superpotential is fixed. Therefore one cannot make the claim that the Hilbert energy-momentum tensor is uniquely specified by ad-hoc adding the divergence of Belinfante superpotential and terms proportional to the equations of motion. In other words, no significant connection exists between Noether's first theorem and the Hilbert energy-momentum tensor in spin-2 Fierz-Pauli theory, as supported by the recent disproof in \cite{baker2021}.

\subsection{6.1) The Belinfante superpotential for spin-2 Fierz-Pauli theory}

We will start by recapping the Belinfante superpotential derivation and showing that it matches the superpotential obtained from the Hilbert energy-momentum tensor in (\ref{FPhilbert}). The Belinfante improvement procedure consists of adding the divergence of a superpotential $b^{[\mu\alpha]\nu}$ to the canonical Noether energy-momentum tensor. By adding this ``improvement'' term ($\partial_\alpha b^{[\mu\alpha]\nu}$) one obtains the Belinfante energy-momentum tensor \cite{belinfante1940}, 

\begin{equation}
T^{\rho\sigma}_B = T^{\rho\sigma}_C + \partial_\alpha b^{[\rho\alpha]\sigma} . \label{belinf}
\end{equation}

The superpotential $b^{[\rho\gamma]\sigma}$ is specifically a combination of the canonical spin angular momentum tensors $S^{\rho [\sigma \gamma]}$ \cite{belinfante1940},

\begin{equation}
b^{[\rho\gamma]\sigma} =
\frac{1}{2} (- S^{\rho [\sigma \gamma]}+S^{\gamma [\sigma \rho]}+S^{\sigma [\gamma \rho]}) , \label{belsuper}
\end{equation}

\noindent a result Belinfante attributes to a Dr. Podolansky (without reference) in his article. The $S^{\rho [\sigma \gamma]}$ are sometimes referred to as the spin contributions. The Belinfante prescription for gravity models has been worked out in \cite{jackiw1994}. The spin angular momentum connection for a second rank $h_{\mu\nu}$ is given by \cite{leclerc2006b},

\begin{equation}
S^{\sigma [\rho\gamma]}=\frac{\partial \mathcal{L}}{\partial \partial_{\sigma} h^{\mu\nu}} [\eta^{\rho \mu} h^{\gamma \nu}-\eta^{\gamma \mu} h^{\rho \nu}+\eta^{\rho \nu} h^{\gamma \mu}-\eta^{\gamma \nu} h^{\rho \mu}] . \label{spinang}
\end{equation}

\noindent Thus we require the derivative of the Fierz-Pauli Lagrangian density  $\frac{\partial \mathcal{L_{FP}}}{\partial (\partial_\sigma h^{\mu\nu})}$ in (\ref{FPL}). Substituting $\frac{\partial \mathcal{L_{FP}}}{\partial (\partial_\sigma h^{\mu\nu})}$ into (\ref{spinang}), we have for the Belinfante superpotential in (\ref{belsuper}),

\begin{multline}
  \hspace*{-1cm}
b^{[\rho\gamma]\sigma} = 
- \frac{1}{2} [\eta^{\gamma\alpha} h^{\sigma \rho}  
- \eta^{\rho\alpha} h^{\gamma \sigma}  ]\partial_{\alpha} h_{\zeta}^{\zeta}
- \frac{1}{2}  [\eta^{\sigma \gamma} h^{\rho \nu} 
-  \eta^{\sigma \rho} h^{\gamma \nu}] \partial_{\nu} h_{\zeta}^{\zeta} 
- [\eta^{\rho \alpha} h^{\gamma \nu} 
-  \eta^{\gamma \alpha} h^{\rho \nu}]  \partial_{\nu} h_\alpha^{\sigma}
- h^{\sigma \nu}  [\partial^{\rho} h_{\ \nu}^{\gamma}
 -   \partial^{\gamma} h_{\nu}^{\rho}] , \label{belinsuperderived}
\end{multline}

\noindent which is exactly what we found for the Hilbert energy-momentum tensor in (\ref{conversesuperpotential}). Thus we have the well known result,

\begin{equation}
b^{[\rho\gamma]\sigma} = \Psi^{[\rho\gamma]\sigma}_H .
\end{equation}

\noindent However, the Belinfante superpotential is not enough by itself to specify the Hilbert tensor (\ref{converseresult}),

\begin{equation}
T^{\rho\sigma}_H = T^{\rho\sigma}_C + \partial_\gamma b^{[\rho\gamma]\sigma} - 2 h^{\sigma}_{\beta} E^{\rho\beta} . \label{hilbeli}
\end{equation}

\noindent We also need the very specific $- 2 h^{\sigma}_{\beta} E^{\rho\beta}$ piece to reconcile the two results; this does not prove uniqueness of $T^{\rho\sigma}_H$ for the Belinfante superpotential. Fixing coefficients in (\ref{genT}) such that the only solutions in Appendix A are those with the specific Belinfante superpotential in (\ref{belinsuperderived}), we will see that the Belinfante superpotential alone yields infinitely many possible results, of which one happens to be the Hilbert expression.

\subsection{6.2) There are infinitely many solutions following from the ad-hoc addition of the divergence of the Belinfante superpotential}

We now prove our main result, that there are infinitely solutions even when fixing the Belinfante superpotential. If we fix (\ref{genadhoc}) with the Belinfante superpotential (\ref{belinsuperderived}) we have,

\begin{equation}
T^{\rho\sigma} = T^{\rho\sigma}_C + \partial_\alpha b^{[\rho\alpha]\sigma}
+ \zeta_1 h E^{\rho\sigma}
+ \zeta_2 h^{\rho}_\alpha E^{\sigma\alpha}
+ \zeta_3 h^{\rho\sigma} {\bf{E}}
+ \zeta_4 h^{\sigma}_{\alpha} E^{\rho\alpha}
+ \zeta_5 \eta^{\rho\sigma} h {\bf{E}}
+ \zeta_6 \eta^{\rho\sigma} h_{\alpha\beta} E^{\alpha\beta} .
\label{genadhocbelin}
\end{equation}

\noindent Therefore if we improve the canonical Noether expression with the divergence of the Belinfante superpotential, we in theory can have the six addition equation of motion pieces. However fixing the Belinfante superpotential coefficients in highly restrictive on the linear system in Appendix A. In particular the superpotential conditions in (\ref{canonsuper1}) to (\ref{canonsuper12}) are much more restricted as now:

\begin{tabular}{@{}p{.5\linewidth}@{}p{.5\linewidth}@{}}
\begin{gather}
B_2 + \bar{D}_1 - \bar{D}_3 = -\frac{1}{2}
\\
b_{9_{i}} = \frac{1}{2}
\\
\bar{B}_2 - \bar{D}_1 + \bar{D}_3 = 0
\\
b_{8_{i}} = 0
\\
b_1 - d_4 = 0
\\
b_3 = 0
\\
d_4 = - 1
\\
b_5 = 1
\\
d_{12_{i}} = 0
\\
B_4 +  \bar{D}_{5_{i}} - \bar{D}_{5_{ii}} = 0
\\
d_{12_{ii}} = 0
\\
b_{11_{ii}}+1 = 0
\\
\bar{B}_4 -  \bar{D}_{5_{i}} + \bar{D}_{5_{ii}} = 1
\\
b_{11_{i}} = - 1
\end{gather}
&
\begin{gather}
b_6+\frac{1}{2} = 0
\\
c_1-\frac{1}{4} = 0
\\
d_8 = 0
\\
c_2+\frac{1}{4} = 0
\\
b_{8_{ii}}-\frac{1}{2} = - \frac{1}{2}
\\
C_5 + \bar{A}_1 - \bar{A}_2 = \frac{1}{2}
\\
\bar{C}_5 - \bar{A}_1 + \bar{A}_2 = 0
\\
b_{9_{ii}}- \frac{1}{2} = 0
\\
b_{10_{ii}}  - d_{12_{ii}} = 0
\\
c_3 = 0
\end{gather}
\end{tabular}

Using Appendix A with these superpotential conditions, and the symmetry conditions in (\ref{symm1}) and (\ref{symm2}), we find a solution with 3 free parameters $\zeta_1$, $\zeta_3$ and $\zeta_5$. The other $\zeta_n$ are $\zeta_2 = 0$, $\zeta_4 = -2$ and $\zeta_6 = 0$. This solution is $a_2 = \frac{1}{2} \zeta_1 + \zeta_5$, $a_4 = - \frac{1}{2} \zeta_1 - \zeta_5$
$b_1 = - 1$, $b_2 = - \frac{1}{2}$, $b_4 = 1$, $b_5 = 1$, $b_6 =- \frac{1}{2}$, $b_7 = \frac{1}{2}$, $b_{9_{i}} = \frac{1}{2}$, $b_{9_{ii}} =  \frac{1}{2} $, $b_{11_{i}} = - 1$, $b_{11_{ii}} = - 1$, 
$c_1 = \frac{1}{4}$, $c_2 = - \frac{1}{4}$, $c_4 = \frac{1}{2}$
$d_1 = \frac{1}{2} - \zeta_3$, $d_2 = - 1 + \zeta_3$, $d_3 = \frac{1}{2} \zeta_1$, $d_4 = - 1$, $d_{6_{i}} = 1$, $d_{6_{ii}} = 1$, $d_7 = \frac{1}{2} \zeta_1$, $d_{9_{i}} = - \frac{1}{2}$, $d_{9_{ii}} = - \frac{1}{2}$, $d_{11_{i}} = - \frac{1}{2}$, $d_{11_{ii}} = - \frac{1}{2}$. Using (\ref{genT}) we have infinitely many conserved and symmetric energy-momentum tensors $T^{\rho\sigma}_{IB}$,

\begin{multline}
T^{\rho\sigma}_{IB} = 
- \partial_\alpha h^{\rho\sigma} \partial_\beta h^{\alpha\beta}
- \frac{1}{2} \partial_\alpha h^{\rho\sigma} \partial^\alpha h 
+ \partial_\alpha h^{\rho}_{\ \beta} \partial^\alpha h^{\sigma\beta}
+ \partial_\alpha h^{\rho\beta} \partial_\beta h^{\sigma\alpha}
\\
- \frac{1}{2} \partial^\rho h \partial^\sigma h 
+ \frac{1}{2} \partial^\rho h_{\alpha\beta} \partial^\sigma h^{\alpha\beta}
+ \frac{1}{2} \partial^\rho h \partial_\alpha h^{\sigma\alpha} 
+ \frac{1}{2} \partial^\sigma h \partial_\alpha h^{\rho\alpha}
\\
- \partial^\rho h_{\alpha\beta} \partial^\alpha h^{\sigma\beta} 
- \partial^\sigma h_{\alpha\beta} \partial^\alpha h^{\rho\beta}
+ \frac{1}{4} \eta^{\rho\sigma} \partial_\alpha h \partial^\alpha h
- \frac{1}{4} \eta^{\rho\sigma} \partial_\alpha h_{\beta\lambda} \partial^\alpha h^{\beta\lambda}
+ \frac{1}{2} \eta^{\rho\sigma} \partial_\alpha h_{\lambda\beta} \partial^\lambda h^{\alpha\beta}
\\
\\
+ (\frac{1}{2} - \zeta_3) h^{\rho\sigma} \partial_\alpha \partial^\alpha h
+ (-1 + \zeta_3) h^{\rho\sigma} \partial_\alpha \partial_\beta h^{\alpha\beta}
+ \frac{1}{2} \zeta_1 h \partial_\alpha \partial^\alpha h^{\rho\sigma}
- h^{\alpha\beta} \partial_\alpha \partial_\beta h^{\rho\sigma}
\\
+ h^{\rho\alpha} \partial_\alpha \partial_\beta h^{\sigma\beta} 
+ h^{\sigma\alpha} \partial_\alpha \partial_\beta h^{\rho\beta}
+ \frac{1}{2} \zeta_1 h \partial^\rho \partial^\sigma h
- \frac{1}{2} h^{\rho\alpha} \partial^\sigma \partial_\alpha h 
- \frac{1}{2} h^{\sigma\alpha} \partial^\rho \partial_\alpha h
- \frac{1}{2} h \partial^\rho \partial_\alpha h^{\sigma\alpha} 
- \frac{1}{2} h \partial^\sigma \partial_\alpha h^{\rho\alpha}
\\
+ (\frac{1}{2} \zeta_1 + \zeta_5) \eta^{\rho\sigma} h \partial_\alpha \partial_\beta h^{\alpha\beta}
+ (- \frac{1}{2} \zeta_1 - \zeta_5) \eta^{\rho\sigma} h \partial_\alpha \partial^\alpha h ,
\end{multline}

\noindent where the subscript $IB$ denotes the `infinite Belinfante' expressions, since using (\ref{genadhoc}) we have infinitely many symmetric, conserved energy-momentum tensors that can be derived from ad-hoc addition of the Belinfante superpotential to the canonical Noether energy-momentum tensor of spin-2 Fierz-Pauli theory,

\begin{equation}
T^{\rho\sigma}_{IB} = T^{\rho\sigma}_C
+ \partial_\alpha b^{[\rho\alpha]\sigma}
+ \zeta_1 h E^{\rho\sigma}
+ \zeta_3 h^{\rho\sigma} {\bf{E}}
- 2 h^{\sigma}_{\alpha} E^{\rho\alpha}
+ \zeta_5 \eta^{\rho\sigma} h {\bf{E}} .
\end{equation}

\noindent Therefore we have proven that adding the ``improvement'' terms associated to the Belinfante superpotential does not specify a unique result; no meaningful connection to Noether's first theorem can be claimed by specifying Belinfante ``improvements''. Note that we solved for particular free coefficients such that we trivially recover the Hilbert energy-momentum tensor (\ref{FPhilbert}) when the free coefficients $\zeta_1$, $\zeta_3$ and $\zeta_5$ are set to zero.

\section{7) Summary and Discussion}

In this article, we considered the most general possible linearized gravity energy-momentum tensor using a procedure developed by Fock \cite{fock2015,infeld1964} and recently applied in a more restricted case to the non-uniqueness problem in linearized gravity by Bi{\v{c}}{\'a}k and Schmidt \cite{bicak2016}. Using this general expression we derived the most general possible superpotential and terms proportional to the equations of motion (\ref{genTsuper}) and used this expression to derive the most general possible ``improvements'' (\ref{genadhoc}) of the canonical Noether energy-momentum tensor of spin-2 Fierz-Pauli theory (\ref{FPcanon}). In Appendix A we gave the most general linear system of equations that represents all such solutions to (\ref{genadhoc}). In addition conditions (\ref{symm1}) and (\ref{symm2}) can be imposed to guarantee symmetry ($T^{\mu\nu} = T^{\nu\mu}$) of the solution. 

Solving this general system in (\ref{genadhoc}) and Appendix A we have proven several results related to the ad-hoc ``improvement'' of energy-momentum tensors in linearized gravity. The addition of a superpotential and terms proportional to the equations of motion to the canonical Noether energy-momentum tensor is often presented as a method for obtaining various energy-momentum tensors from Noether's first theorem, such as the Hilbert (\ref{FPhilbert}) and Landau-Lifshitz (\ref{LLEMT}) expressions in linearized gravity. We have shown that these ad-hoc ``improvements'' do not provide a unique and/or meaningful connection to Noether's first theorem. To highlight this point we derived two new energy-momentum tensors, the Audrey (\ref{AudreyT}) and Elizabeth (\ref{ElizabethT}) energy-momentum tensors. The Elizabeth energy-momentum tensor gives a symmetric expression connected to the canonical Noether tensor in the same way as the Hilbert and Landau-Lifshitz energy-momentum tensors. The Audrey energy-momentum tensor gives a non-symmetric expression, and proves that no symmetric expression can be built when conditions $a_n =0$ and $b_n = 0$ are imposed on (\ref{genT}). Finally we prove that there are infinitely many symmetric and conserved energy-momentum tensors associated to the Belinfante superpotential, one of which is the Hilbert energy-momentum tensors. This is contrary to the conventional wisdom that this association, given in (\ref{hilbeli}), is unique for linearized gravity.

Our results show that there is no unique or meaningful connection between the canonical Noether energy-momentum tensor and any expression obtained after ad-hoc adding the divergence of superpotentials and terms proportional to the equations of motion (i.e. any expressions not derived directly from Noether's first theorem). There are infinitely many such results of this form, and infinitely many even if we restrict our attention to the Belinfante superpotential alone. Selecting a unique energy-momentum tensor for linearized gravity is of course difficult because none are invariant under the spin-2 gauge transformation (linearized diffeomorphisms) as shown by Magnano and Sokolowski \cite{magnano2002}. What is for certain, however, is that outside of the canonical Noether expression, any connection to Noether's first theorem of the various energy-momentum tensors in the literature should be revisited. This is especially true given the recent proof that the Noether and Hilbert energy-momentum tensors are not, in general, equivalent \cite{baker2021}. The question still remains as to what is the physical significance of the many published energy-momentum tensors for linearized gravity in the literature, as well as in general relativity \cite{weinberg1972,papapetrou1948,moller1958,bergmann1953}, and gravity theories as a whole \cite{pimentel1989,koivisto2006,multamaki2008,senovilla2014}. Many of the linearized gravity energy-momentum tensors were highlighted by Bi{\v{c}}{\'a}k and Schmidt \cite{bicak2016}, a study which in part stemmed from continued research in linearized gravity by Butcher et al. in the recent literature \cite{butcher2009,butcher2012,butcher2012b,butcher2012c}. In electrodynamics, fundamental equations such as the Lorentz force law and Poynting's theorem are expressed through the uniquely defined physical energy-momentum tensor. In spin-2 Fierz-Pauli theory, writing down analogous laws requires a unique energy-momentum tensor, for which there is still no consensus on which to choose. In addition, the self coupling problem of $h_{\mu\nu} T^{\mu\nu}$ in the Padmanabhan-Deser debate \cite{padmanabhan2008,deser2010} requires a specific expression which the authors could not agree on, further emphasizing the need for a uniquely defined expression. The linearized Landau-Lifshitz energy-momentum pseudotensor has been used to model observations in one (the Hulse-Taylor) binary pulsar system \cite{hulse1975,taylor1982,peters1963,landau1971}. Others make claims in support of other expressions, such as the Hilbert (metric) energy-momentum tensor in Minkowski spacetime to be the truly physical energy-momentum tensor. For these numerous other linearized gravity energy-momentum tensors in the physics literature, however, experimental or observational verification cannot easily be found.

Due to Magnano and Sokolowski's no-go result \cite{magnano2002}, one can consider energy-momentum tensors in higher derivative gravity in order to obtain a spin-2 gauge invariant expression (invariant under linearized diffeomorphisms), such as the variants of the Bel-Robinson tensor \cite{acquaviva2018,gomez2007}, or the linearized Gauss-Bonnet gravity energy-momentum tensor \cite{petrov2011,baker2019,baker2021b}, which are both invariant under the spin-2 gauge transformation (linearized diffeomorphisms). However, since these models require higher derivative actions, they are not connected to spin-2 Fierz-Pauli theory via standard Lagrangian based energy-momentum derivations such as the Noether method or the Hilbert (metric) method. We note that additional insight about linearized gravity can be found through the Hamiltonian approach \cite{green2011,chishtie2013,smolka2018,waluk2019}. Research on gravitational waves from the linearized gravity equations has continued in recent decades \cite{mashhoon2013}; interests that have only been increasing since the LIGO results in 2016 \cite{abbott2016}. In electrodynamics, the radiation equations are developed from the unique energy-momentum tensor of the theory, emphasizing the need to sort out the non-uniqueness problem in linearized gravity. Various energy-momentum tensors have been proposed to model gravitational radiation, such as the Bel-Robinson tensor \cite{dereli2004,goswami2018}, yet only the linearized Landau-Lifshitz energy-momentum pseudotensor has the observational evidence associated to the Hulse-Taylor binary \cite{hulse1975,taylor1982,peters1963,landau1971}. Sorting out which of the many published expressions correspond to physical phenomena is a fundamental problem which can give great insight into the theoretical framework of gravitational energy. With many published gravitational energy-momentum tensors in the literature, it is not clear which (if any, see philosophical debates on this topic \cite{hoefer2000,read2020}) to use to write down a unique set of physical conservation laws for linearized gravity. We hope that our results will help further progress in this direction, and to clarify the relationship of the many published expressions to the canonical Noether energy-momentum tensor.

\section{Acknowledgement}

We are grateful to N. Kiriushcheva, S. Kuzmin and D.G.C. McKeon for numerous discussions and suggestions while preparing this article.

\bibliographystyle{unsrt}
\bibliography{CanonicalSpin2Bib}

\section{Appendix A - General energy-momentum tensor system of linear equations}

The following Appendix includes the system of linear equations corresponding to the most general energy-momentum tensor for linearized gravity (\ref{genT}) under the condition that the most general expression can be derived from the canonical Noether tensor supplemented by the ad-hoc addition of the divergence of a superpotential and terms proportional to the equations of motion given in (\ref{genadhoc}). In addition to the equations given in this Appendix, (\ref{symm1}) and (\ref{symm2}) can be imposed to derive a symmetric energy-momentum tensor.

\subsection{Canonical Noether conditions}

\begin{gather}
b_7 - d_8 = \frac{1}{2}  \label{canoncondi1}
\\
c_4 = \frac{1}{2} \label{canoncondi2}
\\
b_{10_{i}} - d_{12_{i}} = 0 \label{canoncondi3}
\end{gather}

\subsection{Coefficient splitting conditions}

\begin{tabular}{@{}p{.5\linewidth}@{}p{.5\linewidth}@{}}
\begin{gather}
a_1 = A_1 + \bar{A}_1  \label{splitcanoncondi1}
\\
a_2 = A_2 + \bar{A}_2 
\\
b_2 = B_2 + \bar{B}_2 
\\
b_4 = B_4 + \bar{B}_4 
\end{gather}
&
\begin{gather}
(c_5 + \frac{1}{2}) = C_5 + \bar{C}_5
\\
d_1 = D_1 + \bar{D}_1 
\\
d_3 = D_3 + \bar{D}_3 
\\
d_{5_{i}} = D_{5_{i}} + \bar{D}_{5_{i}} 
\\
d_{5_{ii}} = D_{5_{ii}} + \bar{D}_{5_{ii}}  \label{splitcanoncondi9}
\end{gather}
\end{tabular}

\subsection{Superpotential conditions}

\begin{tabular}{@{}p{.5\linewidth}@{}p{.5\linewidth}@{}}
\begin{gather}
B_2 + \bar{D}_1 - \bar{D}_3 + b_{9_{i}} = 0 \label{canonsuper1}
\\
\bar{B}_2 - \bar{D}_1 + \bar{D}_3 + b_{8_{i}} = 0
\\
b_1 - d_4 + b_3 = 0 = 0
\\
C_5 + \bar{A}_1 - \bar{A}_2 + (b_{8_{ii}} - \frac{1}{2}) = 0 
\\
\bar{C}_5 - \bar{A}_1 + \bar{A}_2 + (b_{9_{ii}} - \frac{1}{2}) = 0 
\\
c_3 + b_{10_{ii}}  - d_{12_{ii}} = 0
\end{gather}
&
\begin{gather}
d_4 + b_5 = 0
\\
d_{12_{i}} + B_4 +  \bar{D}_{5_{i}} - \bar{D}_{5_{ii}} = 0
\\
d_{12_{ii}} + (b_{11_{ii}} + 1) = 0 
\\
\bar{B}_4 -  \bar{D}_{5_{i}} + \bar{D}_{5_{ii}} + b_{11_{i}} = 0
\\
(c_1 - \frac{1}{4})  + (b_6 + \frac{1}{2})  = 0 
\\
(c_2 + \frac{1}{4}) + d_8  = 0  \label{canonsuper12}
\end{gather}
\end{tabular}

\subsection{Linear system of equations for the equations of motion}

\begin{tabular}{@{}p{.5\linewidth}@{}p{.5\linewidth}@{}}
\begin{gather}
- \frac{1}{2} \zeta_1 = \bar{M}_4 \label{canoneomcondi1}
\\
\frac{1}{2} \zeta_1 = D_3 - \bar{B}_2 + \bar{D}_1 
\\
\frac{1}{2} \zeta_1 = d_7 - (b_6 + \frac{1}{2}) 
\\
-\frac{1}{2} \zeta_1 = d_{11_{i}} -  b_{8_{i}} 
\\
-\frac{1}{2} \zeta_1 = d_{11_{ii}} - (b_{9_{ii}} - \frac{1}{2})
\\
\frac{1}{2} \zeta_1 = \bar{M}_3
\end{gather}
  &
\begin{gather}
- \frac{1}{2} \zeta_2 = \bar{M}_2 
\\
\frac{1}{2} \zeta_2 = D_{5_{i}} + \bar{D}_{5_{ii}} - B_4 
\\
\frac{1}{2} \zeta_2 = d_{9_{i}}  -  (b_{8_{ii}} - \frac{1}{2})
\\
-\frac{1}{2} \zeta_2 = d_{10_{i}} + d_{12_{ii}} - b_{10_{ii}} - (b_{11_{ii}} + 1) 
\\
-\frac{1}{2} \zeta_2 = d_{6_{i}}  -  b_5 
\\
\frac{1}{2} \zeta_2 = \bar{M}_1 
\end{gather}
\end{tabular}

\begin{tabular}{@{}p{.5\linewidth}@{}p{.5\linewidth}@{}}
\begin{gather}
 \zeta_3 = M_1 
\\
- \zeta_3 = M_2 
\end{gather}
  &
\begin{gather}
 \zeta_5 = M_3
\\
- \zeta_5 =  M_4
\end{gather}
\end{tabular}

\begin{tabular}{@{}p{.5\linewidth}@{}p{.5\linewidth}@{}}
\begin{gather}
- \frac{1}{2} \zeta_4 = \hat{M}_2 
\\
\frac{1}{2} \zeta_4 = D_{5_{ii}} + \bar{D}_{5_{i}} - \bar{B}_4 
\\
\frac{1}{2} \zeta_4 = d_{9_{ii}} -  b_{9_{i}} 
\\
-\frac{1}{2} \zeta_4 = d_{10_{ii}} + d_{12_{i}}  -  b_{10_{i}}  -  b_{11_{i}}
\\
-\frac{1}{2} \zeta_4 = d_{6_{ii}} -  b_3  
\\
\frac{1}{2} \zeta_4 = \hat{M}_1 
\end{gather}
  &
\begin{gather}
- \frac{1}{2} \zeta_6 = \hat{M}_4
\\
\frac{1}{2} \zeta_6 = a_5 - (c_2 + \frac{1}{4}) 
\\
\frac{1}{2} \zeta_6 = A_1 - C_5 + \bar{A}_2 
\\
-\frac{1}{2} \zeta_6 = \frac{1}{2} (a_3 - c_3 ) 
\\
-\frac{1}{2} \zeta_6 = \frac{1}{2} (a_3 - c_3 ) 
\\
\frac{1}{2} \zeta_6 = \hat{M}_3 \label{canoneomcondi28}
\end{gather}
\end{tabular}

\begin{gather}
d_2 + d_4 - b_1 = M_1 + \bar{M}_1 + \hat{M}_1 \label{Mcanon1}
\\
D_1 - B_2 + \bar{D}_3 = M_2 + \bar{M}_2 + \hat{M}_2  \label{Mcanon2}
\\
A_2 - \bar{C}_5 +  \bar{A}_1 = M_3 + \bar{M}_3 + \hat{M}_3  \label{Mcanon3}
\\
a_4 - (c_1 - \frac{1}{4}) = M_4 + \bar{M}_4 + \hat{M}_4  \label{Mcanon4}
\end{gather}

\end{document}